\documentclass[%
 aip,
 jmp,%
 amsmath,amssymb,
 reprint,%
]{revtex4-1}

\usepackage{graphicx}
\usepackage{dcolumn}
\usepackage{bm}
\usepackage{color}
 \usepackage[version=3]{mhchem}
\usepackage{setspace}
\usepackage{rotating}

\usepackage[final]{pdfpages}

\begin{document}


\title[]{Density Matrix Renormalization Group with Efficient Dynamical Electron Correlation Through Range Separation}

\author{Erik Donovan Hedeg{\aa}rd}
\email{erik.hedegard@phys.chem.ethz.ch}
\affiliation{ETH Z{\"u}rich, Laboratorium f{\"u}r Physikalische Chemie, Vladimir-Prelog-Weg 2, CH-8093 Z{\"u}rich, Switzerland}
\author{Stefan Knecht}
\affiliation{ETH Z{\"u}rich, Laboratorium f{\"u}r Physikalische Chemie, Vladimir-Prelog-Weg 2, CH-8093 Z{\"u}rich, Switzerland}
\author{Jesper Skau Kielberg}
\affiliation{Department of Physics, Chemistry and Pharmacy, University of Southern Denmark, Campusvej 55, Odense, Denmark}
\author{Hans J{\o}rgen Aagaard Jensen}
\email{hjj@sdu.dk}
\affiliation{Department of Physics, Chemistry and Pharmacy, University of Southern Denmark, Campusvej 55, Odense, Denmark}
\author{Markus Reiher}
\email{markus.reiher@phys.chem.ethz.ch}
\affiliation{ETH Z{\"u}rich, Laboratorium f{\"u}r Physikalische Chemie, Vladimir-Prelog-Weg 2, CH-8093 Z{\"u}rich, Switzerland}

\date{May 28, 2015}

\begin{abstract}
We present a new hybrid multiconfigurational method based on the concept of range-separation that combines the density matrix renormalization 
group approach with density functional theory. This new method is designed 
for the simultaneous description of dynamical and static electron-correlation effects 
in multiconfigurational electronic structure problems.

\end{abstract}

\maketitle

\section{\label{intro} Introduction}

Molecular systems with close-lying electronic states possess an electronic structure that is dominated by strong static electron correlation. 
Examples are (i) molecules far from their equilibrium structure and (ii) many transition metal complexes, where static correlation is often sizable.
A method tailored to recover static correlation is the complete active space (CAS) \emph{ansatz}\cite{QUA:QUA23107} for the wave function. This \emph{ansatz} 
defines an active space of $n_{\text{act}}$\ electrons in $N_{\text{act}}$\ orbitals, in which the exact solution of the electronic Schr{\"o}dinger 
equation is obtained by considering a full configuration interaction (FCI) expansion of the wave function. Accordingly, the CAS wave function is defined 
by the number of active electrons and orbitals and the active space is 
denoted as CAS($n_{\text{act}}$,$N_{\text{act}}$). 

In the conventional CAS \emph{ansatz}, the construction of a FCI expansion leads to a factorial scaling with respect to increasing values of 
$n_{\text{act}}$\ and $N_{\text{act}}$. As a consequence, active orbital spaces limited to about CAS(18,18) are computationally feasible. 
By contrast, the density matrix renormalization group (DMRG)\cite{white1992b,white1993}\ algorithm, originally developed for interacting spin chains in solid-state 
physics, iteratively converges to the exact solution in a given active orbital space with polynomial rather than factorial cost\cite{schollwock2011}.  
With DMRG, active orbital spaces are accessible that are about five to six times larger than those of standard CASSCF algorithms. 
To exploit this advantage, an increasing number of quantum-chemistry DMRG implementations have emerged since 
the late 1990s\cite{yaron1998,shuai1998,fano98,white1999,whit00,mitrushenkov2001,chan2002b,legeza2003a,moritz2007,rissler2006,mitrushenkov2012,
sharma2012,nakatani2013,zgid2008a,zgid2008b,kurashige2009,kurashige2013,wouters2012,wouters2014,knec2014,keller2014,maquis2014,dresselhaus2014}. 
Methods, with (DMRG-SCF) and without (DMRG-CI) a simultaneous optimization of the orbital basis, were devised
and a few comprehensive reviews are also available\cite{ors_springer,chan2008a,marti2010b,chan2011,marti2011,wouters2014rev,kurashige2014,szalay2014}. 
Yet, even with larger active orbital spaces at hand, essential parts of the remaining dynamical electron correlation cannot be efficiently accounted for within a DMRG framework. 
Multireference CI (MRCI) in its internally contracted form\cite{saitow2013} and multireference perturbation theory (MRPT) approaches\cite{kurashige2014-cupt2,yanai2015} 
have been combined with DMRG. 
Such approaches may be summarized as 'diagonalize-then-pertub' in the state-specific case and as 'diagonalize-then-perturb-then-diagonalize' in the state-average quasi-degenerate 
case\cite{shavitt2002}. The first step aims at the inclusion of static correlation effects in the zeroth-order Hamiltonian while capturing dynamical correlation in subsequent steps. 
Although potentially accurate, such a strategy comes with a caveat. This caveat is rooted in the need for higher-order reduced density matrices ($n$-RDMs, with $n > 2$) of the 
active space DMRG wave function which leads to a considerably steeper scaling of the MRPT or MRCI step 
compared to the preceding DMRG(-SCF) step. A notable exception is the very recent development of a variational perturbation approach 
that exploits the matrix-product structure of the DMRG wave function and 
optimizes the first-order correction to the wave function iteratively by the DMRG protocol\cite{sharma2014}. 

In this paper, we pursue a yet unexplored strategy to effectively treat dynamical electron correlation 
within the DMRG approach. We propose a simultaneous treatment of dynamic and static correlation
using density functional theory (DFT) for the dynamical-correlation part, rather than aiming for a conventional two-step approach. 
As a consequence, the overall scaling cost does not exceed that of the DMRG calculation. This feature will become particularly advantageous when the approach is applied 
to large systems.  
In turn, the coupling of DFT to a CAS-type wave function method provides the required flexibility for cases where static correlation becomes important and where DFT 
is likely to fail\cite{perdew1995,seminario1994,perdew1996a,gunnarson1976}. 
Although hybrid approaches between wave function theory (WFT) and DFT have not been considered for DMRG yet, they 
have already been studied with other wave function methods \cite{savin1995,grimme1999,savinbook,fromager2007,fromager2010,manni2014,fromager2015}. 
While approximate DFT functionals will introduce errors 
on the absolute scale, relative energies for hybrid WFT--DFT approaches can be obtained with good accuracy\cite{marian2008,silvajunior08,Alam201237-srdft-nevpt2-metallophilicity,fromager2013,hedegaard2013a}.
To avoid the double-counting problem of electron-correlation effects,  
we take advantage of the method originally proposed by Savin\cite{savinbook,savin1995}\ which is based on a range separation of the two-electron 
repulsion operator into a short-range and a long-range part. 
The short-range part of the electron interaction is then treated by DFT, while the long-range part is assigned to the WFT approach. 
The resulting hybrid WFT--DFT method is denoted (\textit{long-range}) WFT--\textit{short-range} DFT (in short, WFT--srDFT)\cite{fromager2007}. In this work, we present 
the (long-range) DMRG--short-range DFT variant, abbreviated as DMRG--srDFT. The theory is formulated generally and applies both to traditional CAS configurational 
interaction (CAS-CI) and to DMRG. It further offers a natural extension to excited states. 

Besides the dynamical correlation problem, a critical issue for \emph{all}\ CAS-based methods is 
the composition of the active orbital space \cite{varyazov2011}. It is a well-established procedure\cite{cheung1979,jensen1988_mp2no,pulay1988} 
to make the orbital choice based on natural orbitals (NOs) and their occupation numbers (NOONs). 
In addition, significant deviations ($> \pm 0.02$) of NOONs from the Hartree-Fock limit of 2  (occupied orbitals) and 0 (virtual orbitals) have been considered 
as multireference indicators. 
In the framework of DMRG, two other measures, namely the single-orbital entropy\cite{legeza2003a} and the mutual information\cite{legeza2006,rissler2006},
have become popular \cite{boguslawski14_entanglement_perspective}. These entropy measures 
are calculated from the one-{\it orbital} RDM --- in the latter case, also from the two-{\it orbital} RDM --- and can be exploited to examine the multireference character  
of a wave function and/or trace the amount of static and dynamic electron correlation in an electronic wave function \cite{boguslawski2012b}. 
Being orbital-based measures, these entropy measures share with NOs the appealing feature that they can simultaneously serve as indicators for multireference character 
as well as selection criteria for the active space. 
With our new DMRG--srDFT approach, we will explore the effect of effectively treating dynamical correlation through the short-range DFT functional 
on entropy measures. 

The paper is organized as follows: In Section \ref{theory}, we briefly outline the theory of range-separated hybrid methods 
and discuss the key steps for a combination of short-range DFT with long-range DMRG. 
Section \ref{implementation_details}\ summarizes technical details of the implementation of our DMRG--srDFT approach. 
Computational details are given in Section \ref{Compdetails} before we proceed with the first applications of DMRG--srDFT in Section \ref{results}. 

We discuss and evaluate our approach in practical calculations. For  \ce{H2O} and \ce{N2}, 
we investigate the effect of different active spaces up to FCI for the long-range wave function along the symmetrical bond stretching coordinate.  
As their ground-state electronic structure becomes increasingly multiconfigurational upon bond elongation, 
these molecules can serve as prototypical examples for both singlereference and multireference methods, 
We highlight differences in entropy measures calculated with DMRG and DMRG--srDFT wave functions to show that srDFT produces a stable
pattern of these measures that is rather independent of the size of the active orbital space. 
As a final application we investigate two ligand-dissociation reactions of $d$-block metal complexes taken from the 
WCCR10 benchmark set\cite{weymuth2014,weymuth2015}, for which accurate experimental reference data in the gas phase at zero Kelvin are available.
Section \ref{conclusion} summarizes our findings and outlines future developments.
 
\section{\label{theory}Theory}

In this paper, we generally work in Hartree atomic units and exploit the second-quantization 
formalism. Orbital indices $p,q,r,s$ denote spatial general orbitals, $i,j,k,l$ inactive (doubly occupied) orbitals, and $u,v,x,y$ active (partially occupied) orbitals, thus following the 
notation by Roos, Siegbahn, and co-workers\cite{roos1980b,siegbahn1981}. The electronic non-relativistic Hamiltonian then reads as 
\begin{equation}
 \hat{H} = \sum_{pq}h_{pq}\hat{E}_{pq} + \frac{1}{2}\sum_{pqrs} g_{pqrs}\hat{e}_{pqrs} + \hat{V}_{\text{nn}} \label{hamilton} ,
\end{equation}
where $\hat{V}_{\text{nn}}$ is the nuclear repulsion potential energy operator and the one- and two-electron integrals over 
molecular orbitals $\phi_i (\bm{r})$  are defined as
\begin{align}
  h_{pq}   & = \langle\phi_{p}(\bm{r})\vert\, \hat{h}\, \vert \phi_{q}(\bm{r})\rangle  = \langle p \vert\, \hat{h}\, \vert q \rangle \label{h_pq}  \\[1.0ex]
 g_{pqrs} & = \langle \phi_{p}(\bm{r}_1) \phi_{r}(\bm{r}_2) \vert \hat{g}(1,2) \vert \phi_{q}(\bm{r}_1) \phi_{s}(\bm{r}_2)\rangle \notag \\
 & = \langle p r  \vert \hat{g}(1,2) \vert q  s\rangle . \label{g_pqrs}
\end{align}
The operators $\hat{h}$ and $\hat{g}(1,2)$ are given in first quantization: 
As usual, $\hat{h}$ contains the operators for the kinetic energy of an electron and its interaction with all nuclei in the system, 
whereas $\hat{g}(1,2)$ is the two-electron repulsion operator
\begin{align}
\hat{g}(1,2) = \frac{1}{|\bm{r}_1 - \bm{r}_2 |} . \label{g_ij}
\end{align} 
The $\hat{E}_{pq}$ and $\hat{e}_{pqrs}$ operators are defined in terms of creation and annihilation operators,
\begin{align}
 \hat{E}_{pq}   = \sum_{\sigma}\hat{a}^{\dagger}_{p\sigma}\hat{a}_{q\sigma} &&\mbox{and}&&
 \hat{e}_{pqrs} = \hat{E}_{pq}\hat{E}_{rs} - \hat{E}_{ps}\delta_{qr} . 
\end{align}
Then, the electronic energy for an electronic wave function $\Psi$ can  
be written in terms of the one- 
and two-electron RDMs, 
\begin{align}
&  D_{pq} = \langle \Psi \vert \hat{E}_{pq}\vert \Psi \rangle \label{one-and-two-electron-dens-1} 
\end{align}
and
\begin{align}
&  P_{pqrs} = \langle \Psi \vert \hat{e}_{pqrs}\vert \Psi \rangle \label{one-and-two-electron-dens-2} ,
\end{align}
respectively, as 
\begin{equation}
 E = \langle \Psi \vert \hat{H}\vert \Psi \rangle = \sum_{pq}h_{pq}D_{pq} + \frac{1}{2}\sum_{pqrs}g_{pqrs}P_{pqrs} + {V}_{\text{nn}},
\end{equation}
where the operator $\hat{V}_{\text{nn}}$ is written as a potential energy ${V}_{\text{nn}}$ of the nuclear framework since it does not depend on electronic coordinates,
which are the dynamical variables integrated out in the energy expectation value.
In CAS-type methods, parts of the RDMs will be associated with inactive electrons and the computational evaluation of the energy expression can be further simplified by 
splitting it into separate contributions for inactive and active electrons.
The corresponding formalism will be elaborated in the following subsection.  

\subsection{Complete-Active-Space Configuration Interaction}

By dividing the one-electron RDM into an inactive part ('I'), 
$\bm{D}^{\text{I}} = \{ D^{\text{I}}_{ij}\} = \{ 2 \delta_{ij}\}$, 
and an active part ('A'), $\bm{D}^{\text{A}} = \{ D^{\text{A}}_{uv}\}$,
we may write the CAS-CI energy expression as a sum of an inactive energy ($E_{\rm I}$)
and an active energy ($E_{\rm A}$):
\begin{align}
E_{\text{CAS-CI}} = E_{\text{I}} + E_{\text{A}} , \label{CAS-CI-energy} 
\end{align}
where
\begin{align}
E_{\rm I} & =  \frac{1}{2}\sum_{ij}\bigl(h_{ij} + f^{\text{I}}_{ij}\bigr) D^{\text{I}}_{ij} + V_{\text{nn}} \notag \\
              & = \sum_{i}\bigl(h_{ii} + f^{\text{I}}_{ii}\bigr) + V_{\text{nn}}  \\
E_{\rm A}   & = \sum_{uv}f^{\text{I}}_{uv}D^{\text{A}}_{uv} + \frac{1}{2}\sum_{uvxy}g_{uvxy}P^{\rm A}_{uvxy} \label{act-energy} . 
\end{align} 
The inactive energy $E_{\rm I}$ is equal
to the Hartree-Fock energy expression for the doubly-occupied orbitals. 
The matrix element $f^{\text{I}}_{pq}$ denotes an element of the inactive Fock matrix 
\begin{equation}
 f^{\text{I}}_{pq} = h_{pq} + \sum_{k}\bigl(2 g_{pqkk} - g_{pkqk} \bigr) \label{Fock_inact} ,
\end{equation}
which has been defined according to Eq.~(15a) in Ref.~\citenum{siegbahn1981} (see also Ref.~\citenum{helgaker2004_pages482_and_645-646} for explicit derivations).  
In $E_{\rm A}$, the use of the inactive Fock matrix, $\bm{f}^{\rm I} = \{f^{\rm I}_{uv}\}$, instead of the one-electron matrix, $\bm{h} = \{h_{uv}\}$, accounts for the 
screening of the nuclei by the inactive electrons.

The energy expressions in Eqs.~\eqref{CAS-CI-energy}--\eqref{act-energy} hold for any CAS-type method, including those with orbital optimization such as the 
Complete-Active-Space Self-Consistent-Field 
(CASSCF) method. 

\subsection{Range-separated CAS-CI hybrids with DFT}\label{Kohn-Sham-DFT-multideterminantal}

The two-electron repulsion operator can be separated into a long-range ('lr') part and a short-range ('sr') part\cite{savinbook,fromager2007,fromager2008,fromager2013,hedegaard2013b},
\begin{equation}
\hat{g}(1,2) = \hat{g}^{\mu,\text{lr}}(1,2) + \hat{g}^{\mu,\text{sr}}(1,2) , 
\label{sr-DFTCoulpart}
\end{equation}
involving a range-separation parameter $\mu$.
This decomposition of the electron-electron interaction operator has been applied in various WFT--srDFT 
hybrid methods\cite{savinbook,savin1995,goll2005,fromager2007,fromager2008,JCP12_Pernal_tddmft-srdft,fromager2013}. 
In this paper, the long-range and short-range parts of the interaction operator are separated by virtue of the error function\cite{savin1995},
\begin{align}
  \hat{g}^{\mu,\text{lr}}(1,2) & = \frac{\text{erf} (\mu | \bm{r}_{1} - \bm{r}_{2} |)}{|\bm{r}_{1} - \bm{r}_{2}|}     \label{lrpart} \\[3.0ex]
  \hat{g}^{\mu,\text{sr}}(1,2) & = \frac{1- \text{erf} (\mu | \bm{r}_{1} - \bm{r}_{2} |)}{|\bm{r}_{1} - \bm{r}_{2}|}  \label{sr-DFTCoulpart-2} .
\end{align} 
In the following, the long-range and short-range two-electron integrals, $g^{\text{lr}}_{pqrs}$ and $g^{\text{sr}}_{pqrs}$,
are the integrals in which $\hat{g}(1,2)$ of Eq.~\eqref{g_pqrs} has been replaced by $\hat{g}^{\mu,\text{lr}}(1,2)$ 
and  $\hat{g}^{\mu,\text{sr}}(1,2)$, respectively.
All two-electron integrals depend on the range-separation parameter $\mu$ to be fixed prior to a calculation.
For the sake of brevity, we refrain from denoting this explicit dependency for the integrals and for the energy in what follows.

In the next step, the short-range part of the electron--electron interaction energy is described by DFT with a tailored (short-range) functional $E^{\text{sr}}_{\text{Hxc}}[\rho]$ of the total electron density 
\begin{align}
& \rho(\bm{r}) =  \langle\Psi\vert \hat{\rho}\vert \Psi\rangle = \sum_{pq}\langle\Psi\vert \Omega_{pq}\hat{E}_{pq}\vert \Psi\rangle = \sum_{pq}\Omega_{pq}D_{pq} , \notag \\
&\mbox{with}\quad  \Omega_{pq}(\bm{r}) = \phi^{*}_{p}(\bm{r})\phi_{q}(\bm{r}) \label{rho} .
\end{align}
We note that the limits $\mu=0$ and $\mu\rightarrow\infty$ then correspond to Kohn-Sham DFT and {\it ab initio} WFT, respectively.

The srDFT functional is partitioned as usual in DFT methodology
into a Hartree (Coulomb) term, $E^{\text{sr}}_{\text{H}}[\rho]$, and an exchange-correlation (xc) contribution, $ E^{\text{sr}}_{\text{xc}}[\rho]$,
\begin{equation}
 E^{\text{sr}}_{\text{Hxc}}[\rho] =  E^{\text{sr}}_{\text{H}}[\rho] + E^{\text{sr}}_{\text{xc}}[\rho] \label{E_Hxc_srDFT},
\end{equation}
where
\begin{equation}
 E^{\text{sr}}_{\text{H}}[\rho] = \dfrac{1}{2} \sum_{pq,rs} D_{pq}\, g^{\text{sr}}_{pqrs}\, D_{rs} = 
 \dfrac{1}{2} \sum_{pq} j^{\text{sr}}_{pq}\, D_{pq} , \label{E_H_energy}
\end{equation}
while the explicit form of $E^{\text{sr}}_{\text{xc}}[\rho]$ depends
on the choice of the approximate functional. We have implicitly defined the short-range two-electron Coulomb potentials $j^{\text{sr}}_{pq}$ in Eq.~\eqref{E_H_energy}. 

It must be stressed that results for a range-separated hybrid WFT--DFT approach in practice   
will be $\mu$-dependent due to the approximate nature of the short-range functionals available. 
The same is true for range-separated Kohn-Sham DFT. 
Calibration studies for the latter suggest that values in the interval 0.33 a.u. $<$ $\mu$ $<$ 0.5 a.u. 
are optimal\cite{ikura2001,tawada2004,vydrov2006a,vydrov2006b,gerber2005}. Studies using a CAS--srDFT hybrid\cite{fromager2007} have shown 
that $\mu=0.4$ a.u. is a good compromise which optimizes the amount of static correlation recovered by the wave function part. 
An alternative which defined the optimal $\mu$-value as the value that provides the lowest CAS--srDFT energy was also explored\cite{fromager2007}, 
but was found to be system dependent (due to the approximate srDFT functional). It is therefore not surprising that 
for some systems the lowest energy can be obtained in either pure CASSCF ($\mu = \infty$) or pure DFT ($\mu=0$) calculations.

For finite $\mu$, the CAS-CI energy expression of Eq.~\eqref{CAS-CI-energy} becomes
\begin{equation}
E_{\text{CAS-CI}}^{\text{srDFT}} = E^{\text{lr}}_{\text{I}} + E^{\text{lr}}_{\text{A}} 
+  E^{\text{sr}}_{\text{H}}[\rho] + E^{\text{sr}}_{\text{xc}}[\rho] \label{CAS-CI-srDFT-energy} ,  
\end{equation}
where the first two terms are identical to Eq.~\eqref{CAS-CI-energy} except that all regular two-electron integrals
have been replaced by the long-range two-electron integrals, that is, $g_{pqrs}\rightarrow g^{\text{lr}}_{pqrs}$. Accordingly, the 
inactive Fock matrix in Eq.~\eqref{Fock_inact} is modified to
\begin{equation}
 f^{\text{I},\text{lr}}_{pq} = h_{pq} + \sum_{k}\bigl(2 g^{\text{lr}}_{pqkk} - g^{\text{lr}}_{pkqk} \bigr) . 
\end{equation}
One notes that this CAS-CI--srDFT energy expression is not linear in the one- and two-electron density matrices as in standard CAS-CI, 
because the Hartree and exchange-correlation terms in Eq.~\eqref{CAS-CI-srDFT-energy} are non-linear in the one-electron density matrix. We illustrate this by  
considering a linear deviation $\Delta D_{pq} =  D_{pq} - D^{\text{ref}}_{pq} $ 
from some (fixed) reference density matrix, $\bm{D}^{\text{ref}} = \{ D^{\text{ref}}_{pq}\}$. 
The one-electron density matrix elements are thus 
\begin{equation} 
D_{pq} = D^{\text{ref}}_{pq} + \Delta D_{pq} , 
\end{equation}
which by insertion in Eq.~\eqref{rho} leads to
\begin{equation}
\rho = \rho^{\text{ref}}+\Delta\rho , 
\end{equation} 
in an obvious notation.
As  $E^{\text{sr}}_{\text{Hxc}}[\rho]$  is non-linear in the one-electron density matrix, we note that
\begin{equation}
E^{\text{sr}}_{\text{Hxc}}[\rho^{\text{ref}}\!+\!\Delta\rho] \neq E^{\text{sr}}_{\text{Hxc}}[\rho^{\text{ref}}] + E^{\text{sr}}_{\text{Hxc}}[\Delta\rho] . 
\end{equation}
This has the consequence that an exact CAS-CI--srDFT expression is \textit{state specific}, 
and we cannot diagonalize a matrix to obtain exact CAS-CI--srDFT electronic energies of several roots.
As holds in general for state-specific methods, this implies that the CI expansions for different states of the same symmetry will be non-orthogonal.

However, following Pedersen\cite{pedersenphd2004}, we can define a linear model providing orthogonal CI states
in the spirit of \textit{state-averaged} CASSCF
by using the following linear approximation to the energy change, 
\begin{align}
	E^{\text{sr}}_{\text{Hxc}}[\rho^{\text{ref}}&\!+\!\Delta\rho] -
    	E^{\text{sr}}_{\text{Hxc}}[\rho^{\text{ref}}] \\ \nonumber
    	\approx& \int \,\frac{\delta E^{\text{sr}}_{\text{Hxc}}}{\delta \rho(\bm{r})}[\rho^{\text{ref}}]\, \Delta \rho(\bm{r})\, \text{d}\bm{r} \\ \nonumber
        =& \sum_{pq} (j^{\text{ref,sr}}_{pq} + v^{\text{ref,sr}}_{pq} ) \Delta D_{pq} ,
\end{align}
where
\begin{align}
 j^{\text{ref,sr}}_{pq} =& \langle\phi_{p}\vert \hat{j}^{\text{sr}}_{\text{xc}}[\rho^{\text{ref}}] \vert\phi_{q}\rangle 
                        = \langle\phi_{p}\vert \frac{\delta E^{\text{sr}}_{\text{H}}}{\delta \rho(\bm{r})}[\rho^{\text{ref}}] \vert\phi_{q}\rangle \\ \nonumber
 =& \sum_{rs} g^{\text{sr}}_{pqrs}\, D^{\text{ref}}_{rs}  ,
  \label{j_integrals}
\end{align}
are the matrix elements of the short-range Coulomb operator, and
\begin{equation}
 v^{\text{ref,sr}}_{pq} = \langle\phi_{p}\vert \hat{v}^{\text{sr}}_{\text{xc}}[\rho^{\text{ref}}]  \vert\phi_{q}\rangle
                        = \langle\phi_{p}\vert \frac{\delta E^{\text{sr}}_{\text{xc}} }{\delta \rho(\bm{r})}[\rho^{\text{ref}}] \vert\phi_{q}\rangle , 
 \label{v_xc_integrals}
\end{equation}
are the matrix elements of the short-range exchange-correlation potential.
We can now define the state-averaged CAS-CI--srDFT method for $M$ electronic states, using the reference density matrix elements
\begin{equation}
D^{\text{ref}}_{pq} = \frac{1}{M}\sum_{i=1}^{M} w_i D^{i}_{pq} ,
\end{equation}
and thus $\rho^{\text{ref}} = \frac{1}{M} \sum_{i=1}^{M} w_i \rho_i$, where the weights $w_i$ add up to one.
Although the CI expansions now will be orthogonal, the equations are still non-linear.
The most straight forward optimization procedure is an iterative method, in many ways similar to Hartree-Fock theory;
this will be described in Sec.~\ref{implementation_details}.

We proceed by noting that in any CAS-CI model $\Delta D_{pq} = \Delta D^{\text{A}}_{uv}$ because the inactive part, $D^{\text{I}}_{ij}$, is
fixed by definition. The state-averaged CAS-CI--srDFT energy for the $i=1,M$ selected roots can then be written as
\begin{eqnarray}
E_{\text{CAS-CI}}^{\text{srDFT},i} &=& E^{\text{lr}}_{\text{I}} + E^{\text{lr},i}_{\text{A}}   
 +  E^{\text{sr}}_{\text{Hxc}}[\rho^{\text{ref}}] \nonumber \\ 
        && + \sum_{uv} ( j^{\text{ref,sr}}_{uv} + v^{\text{ref,sr}}_{uv} ) \Delta D^{\text{A},i}_{uv}  . 
\label{CAS-CI-SA-srDFT-energy}
\end{eqnarray}
If only one root is used ($M=1$), the formalism will coincide with a state-specific optimization. 
In the DMRG--srDFT variant, all equations above hold, but a DMRG protocol (see the following subsection) optimizes the active density matrices.

\subsection{The DMRG ansatz and correlation measures}

A given electronic wave function can be expanded in terms of occupation number vectors,   
\begin{equation}
 \Psi = \sum_{\sigma_{1}\cdots\sigma_{L}} \psi_{\sigma_{1}\cdots\sigma_{L}}\vert\sigma_{1}\cdots\sigma_{L} \rangle . \label{represnetation_form}
\end{equation}
Here, the configuration-interaction expansion coefficients are written as a \textit{coefficient tensor}, $\psi_{\sigma_{1}\cdots\sigma_{L}}$, according to the
tensorial construction of the $4^{N_{\rm act}}$-dimensional Hilbert space from $N_{\rm act}$ spatial orbitals.  
In DMRG terminology, each spatial orbital defines a \textit{site} with four possible one-electron states,
\begin{equation}
 \sigma_{j} = \left\{ \vert vac\rangle, \vert \alpha\rangle, \vert \beta\rangle, \vert\alpha\beta\rangle \right\}_j .  
\end{equation} 
The quantum-chemical DMRG approach builds up the CAS wave function by first arranging the set of (active) orbitals $ \{\phi^{\rm A}_u \}$ in a linear 
order according to some optimization recipe (e.g., according to the single-orbital entropies calculated from a few DMRG sweeps).
In our second-generation, i.e., matrix-product-operator-based implementation of the DMRG algorithm\cite{keller2014,maquis2014},
each site has an associated set of operators in matrix representation. While optimizing the site matrices iteratively, 
the DMRG protocol constitutes a variational optimization with respect to the total electronic energy.
One- and two-electron RDMs as in Eqs.~\eqref{one-and-two-electron-dens-1} and \eqref{one-and-two-electron-dens-2}, can then be evaluated. 
For our DMRG--srDFT method, these density matrices are required to evaluate the energy expression in 
Eq.~\eqref{CAS-CI-SA-srDFT-energy}. 

In order to estimate the multireference character of the target molecule in terms of orbital-based 
descriptors, we exploit the fact that the DMRG wave function can be easily partitioned into two (open) quantum systems within the DMRG algorithm: 
One or two orbitals are embedded into all remaining orbitals of the active space.  
If $\vert n\rangle$ denotes the states defined on this single orbital (or on the two orbitals, respectively) and $\vert j \rangle$ those defined on the remaining orbitals of the CAS,
the partitioning yields an RDM operator for the states defined on the embedded orbital(s), 
\begin{align}
 \hat{\rho}_{nn'}  = \sum_{jj'}\vert j \rangle \vert n \rangle \langle n' \vert \langle j' \vert \label{red_den_operator_1}   , 
\end{align}
where the \textit{environment states} $\vert j \rangle$ and $\vert j' \rangle$ are traced out.
The one-orbital (or two-orbital) RDM evaluated as an expectation value of the RDM operator in Eq.~\eqref{red_den_operator_1},  
\begin{align}
 \rho_{nn'} & = \langle \Psi\vert \hat{\rho}_{nn'}\vert \Psi\rangle , \label{1-rdm}
\end{align}
is then diagonalized to obtain four (or sixteen) eigenvalues $n_{\alpha,u}$ (or $n_{\alpha,uv}$) for orbital $\phi^{\rm A}_u$ (or for orbitals $\phi^{\rm A}_u$ and $\phi^{\rm A}_v$). 
The one-orbital entropy\cite{legeza2003a},
\begin{equation}
s_{u} = -\sum^{4}_{\alpha=1}n_{\alpha,u}\ln(n_{\alpha,u})  , \label{one-electron-entropy}
\end{equation}
measures the degree of \textit{entanglement}\cite{huang2005} 
of the four possible states defined on orbital $u$ with all states defined on the environment orbitals. 
Similarly to Eq.~\eqref{one-electron-entropy}, the entanglement of all 16 states defined on two orbitals embedded in the complementary orbital space of the CAS is measured by 
the two-orbital entropy,
\begin{equation}
s_{uv} = -\sum^{16}_{\alpha=1}n_{\alpha,uv}\ln(n_{\alpha,uv}) . \label{two-electron-entropy}
\end{equation}
Eq.\ \eqref{two-electron-entropy} also contains single-orbital contributions, which can be eliminated
by subtracting the single-orbital entropies of orbitals $u$ and $v$, which defines the
mutual information\cite{legeza2003a,legeza2006,rissler2006},
\begin{equation}
 I_{uv} = \frac{1}{2}\left(s_{uv} - s_{u} - s_{v}\right)\left(1-\delta_{uv}\right) . 
\end{equation}

\section{Implementation}\label{implementation_details}

Our DMRG--srDFT implementation is based on an existing CI--srDFT program\cite{pedersenphd2004}\ and provides an interface between a 
development version of the {\sc Dalton}\cite{DALTON2015, WCMS:WCMS1172}\ program and the {\sc Maquis} quantum chemical DMRG program\cite{maquis2014}. 
{\sc Maquis} is a genuine DMRG program based on matrix product states (MPS) and matrix product operators (MPO). 
Both regular and short-range integrals are calculated with {\sc Dalton}. 
\begin{figure}[t!]
 \centering
 \includegraphics[height=5.8cm]{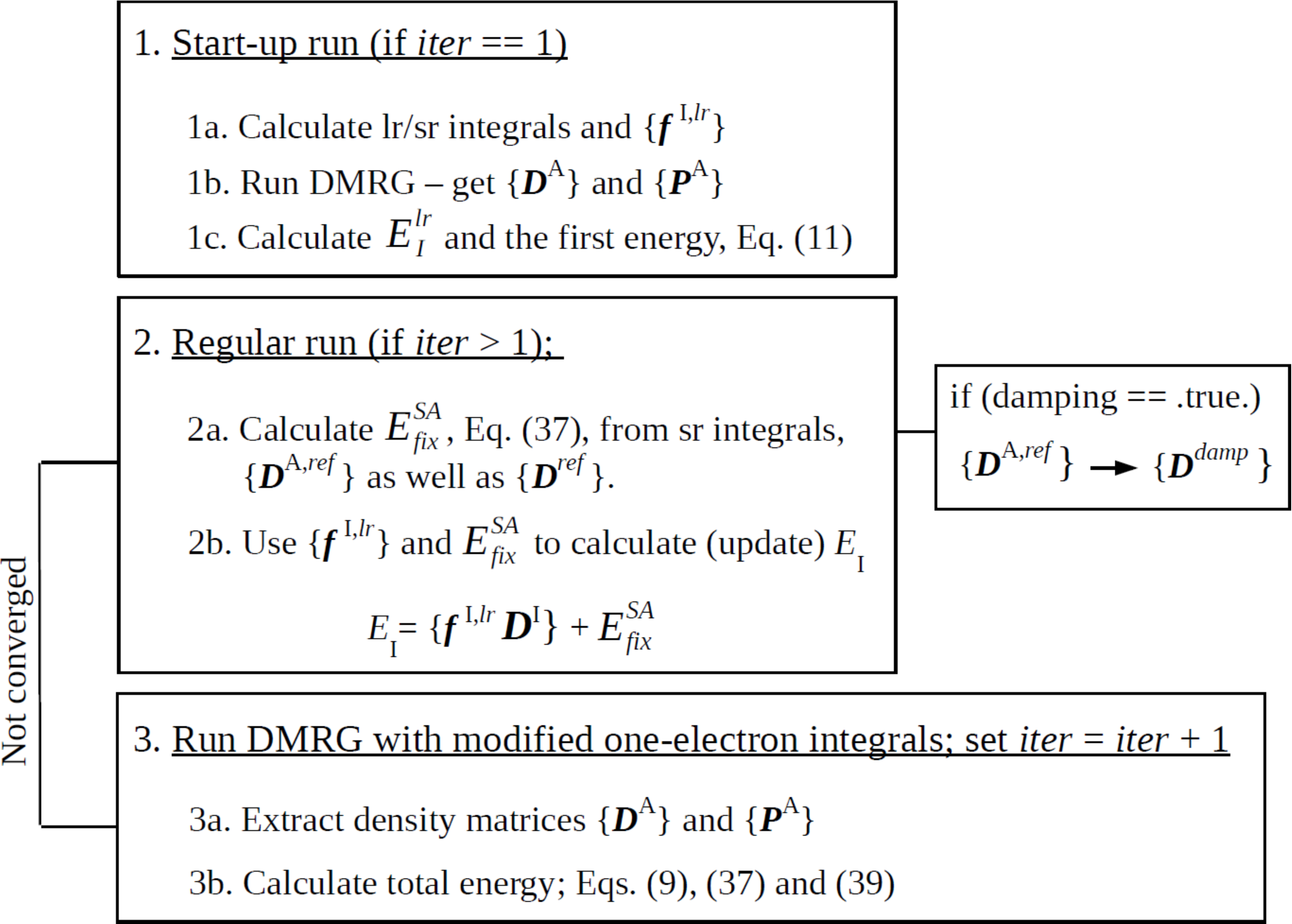}
 \caption{Flow chart for the DMRG--srDFT implementation reported in this work. Each $iter$ denotes a macro-iteration, and the DMRG (or CI) uses sufficient sweeps 
(or micro-iterations) to achieve convergence within each macro-iteration. \label{dmrg_implementation}}
\end{figure}
To be closer to the operational expressions used in the program we split the total energy in Eq.~\eqref{CAS-CI-SA-srDFT-energy} into  inactive and active parts as in Eq.~\eqref{CAS-CI-energy}. 
The inactive  part then becomes  
\begin{align}
 E^{\text{srDFT}}_{\rm I}  &=   E^{\text{lr}}_{\rm I} +   E^{\rm SA}_{\rm fixed},  
 \label{E_inact_final} 
\end{align} 
where the \textit{fixed state-average} ('SA') energy, $E^{\rm SA}_{\rm fixed}$, contains short-range terms that depend only on the inactive and the reference density matrices 
\begin{align}
E^{\rm SA}_{\rm fixed} & =  \frac{1}{2}\sum_{ij} j^{\text{I},\text{sr}}_{ij}  D^{\text{I}}_{ij} -\frac{1}{2}\sum_{uv} j^{\text{A},\text{ref,sr}}_{uv} D^{\text{A},\text{ref}}_{uv}  \notag \\
 +& E^{\rm sr}_{\rm xc}[\rho^{\rm ref}]
-  \sum_{uv}v^{\text{ref},\text{sr}}_{\text{xc},uv}D^{\text{A},\text{ref}}_{uv} , 
\end{align}
with the state-averaged active reference density matrix 
\begin{equation}
D^{\text{A},\text{ref}}_{uv} = \frac{1}{M}\sum_{i=1}^{M} w_i D^{\text{A},i}_{uv} . 
\end{equation}
Recall that the active reference density matrix is kept fixed during the CI or DMRG optimization, and can therefore  
be assigned to the inactive energy. It changes only in each macro-iteration as it is obtained from the wave function of 
the previous iteration.
The active part of the energy is 
\begin{align}
 E^{\text{srDFT},i}_{\text{A}}  & = E^{\text{lr},i}_{\text{act}} 
+ \sum_{uv}\bigl(j^{\text{I},\text{sr}}_{uv} + j^{\text{A},\text{ref,sr}}_{uv}  + v^{\text{ref},\text{sr}}_{\text{xc},uv} \bigr) D^{\text{A},i}_{uv} \notag \\
 & = \sum_{uv}\bigl(f^{\text{I},\text{lr}}_{uv} + j^{\text{I},\text{sr}}_{uv} + j^{\text{A},\text{ref,sr}}_{uv}  + v^{\text{ref},\text{sr}}_{\text{xc},uv} \bigr) D^{\text{A},i}_{uv}  \notag \\ 
+ &  \frac{1}{2}\ \sum_{uvxy}g^{\text{lr}}_{uvxy}P^{\text{A},i}_{uvxy}  
\label{E_act_final}  .
\end{align}   
We can now define the iterative CAS-CI--srDFT or DMRG--srDFT procedure as follows:
\begin{enumerate}
 \item select an initial reference density;
 \item solve for the $M$ CI vectors / DMRG states used in the averaging with fixed reference density;
 \item calculate new reference density and energy; if not converged, go back to step 2.
\end{enumerate}
Figure \ref{dmrg_implementation}\ summarizes the workflow in steps 1--3 of our implementation.
Note that integrals with four active indices are constant during the iterations. 
Thus, they are calculated (and transformed to the MO basis) only once. Moreover, $E^{\rm lr}_{\rm I}$ is also a constant. 
The only quantities that need to be recalculated in each (macro)-iteration are $j^{\text{A},\text{ref,sr}}_{uv}$ and
$v^{\text{ref},\text{sr}}_{\text{xc},uv}$. 

We emphasize that steps 1--3 and the scheme in Figure \ref{dmrg_implementation} hold for both DMRG--srDFT and general CI--srDFT (if DMRG is replaced by CI). 
The first-order optimizer in the previous work\cite{pedersenphd2004} did not consider any convergence acceleration or damping schemes and convergence problems were 
frequently observed. In order to (partially) solve the latter issue we introduce a simple, dynamical damping scheme. In iteration $iter$, we modify $\bm{D}^{\text{A},\text{ref}}$ 
according to
\begin{equation}
 \bm{D}^{\text{damp}}_{iter} = \alpha\bm{D}^{\text{A},\text{ref}}_{iter} +  (1-\alpha)\bm{D}^{\text{A},\text{ref}}_{iter-1} , 
\end{equation}
where $\alpha$ is a dynamically adjusted damping factor. 

\section{\label{Compdetails} Computational Details}       
  
Calculations for \ce{H2O} and \ce{N2} employed a Dunning cc-pVDZ basis set\cite{bs890115d} for O, H, and N. 
The DMRG--srDFT calculations were in all cases performed with the srPBE functional from Ref.~\citenum{fromager2007} i.e.~with the 
Heyd--Scuseria--Ernzerhof\cite{heyd2003} short-range exchange functional together with the rational interpolation srPBE correlation functional 
defined in Ref.~\citenum{fromager2007} (SRCPBERI in DALTON). 
The range-separation parameter  was always set to $\mu = 0.4$ a.u.\cite{fromager2007}. 

The structures considered in this work correspond to the ones used by  
Olsen \textit{et al.}\cite{olsen1996b} for \ce{H2O} and by Chan \textit{et al.}\cite{chan2004b} for \ce{N2}. For \ce{H2O}, we exclusively consider 
the symmetric stretch coordinate. 
The truncated active orbital spaces for \ce{H2O} and \ce{N2} comprise all valence electrons and orbitals required for a balanced description 
of the valence electronic structure along the stretching mode, that is DMRG(8,8) for \ce{H2O} (corresponding to 'CASB' in Ref.\citenum{olsen1996b}) and DMRG(6,6) for \ce{N2}.
\ce{H2O} was calculated in $\text{C}_{2\text{v}}$ symmetry and the CAS(8,8) space includes 4 orbitals in $\text{A}_{1}$ symmetry, 
and 2 orbitals in $\text{B}_{1}$ and $\text{B}_{2}$ symmetries, respectively. 
\ce{N2} was calculated in the $\text{D}_{2\text{h}}$ subgroup. Thus, the  used CAS(6,6) space correspond to 1 orbital in symmetries  $\text{A}_{\text{g}}$, 
 $\text{B}_{3\text{u}}$,  $\text{B}_{2\text{u}}$,  $\text{B}_{1\text{u}}$,  $\text{B}_{2\text{g}}$ and  $\text{B}_{3\text{g}}$.  
Active spaces corresponding to FCI (DMRG-FCI) within the cc-pVDZ basis set are DMRG(10,24) for \ce{H2O} and DMRG(14,28) for \ce{N2}. 
The number of renormalized DMRG block states $m$ was set to $m=512$\ for \ce{H2O} while for \ce{N2} higher $m$\ values of up to  
$m=2048$\ were required for technical reasons to achieve similar convergence as in Ref. \citenum{chan2004b}. 
Accordingly, we specify the DMRG data as DMRG($n_{\text{act}}$,$N_\text{act}$)[$m$].
All calculations have been performed as state-specific ones, i.e., with one root.

\begin{figure}[htb!]
  \centering
  \includegraphics[height=5cm]{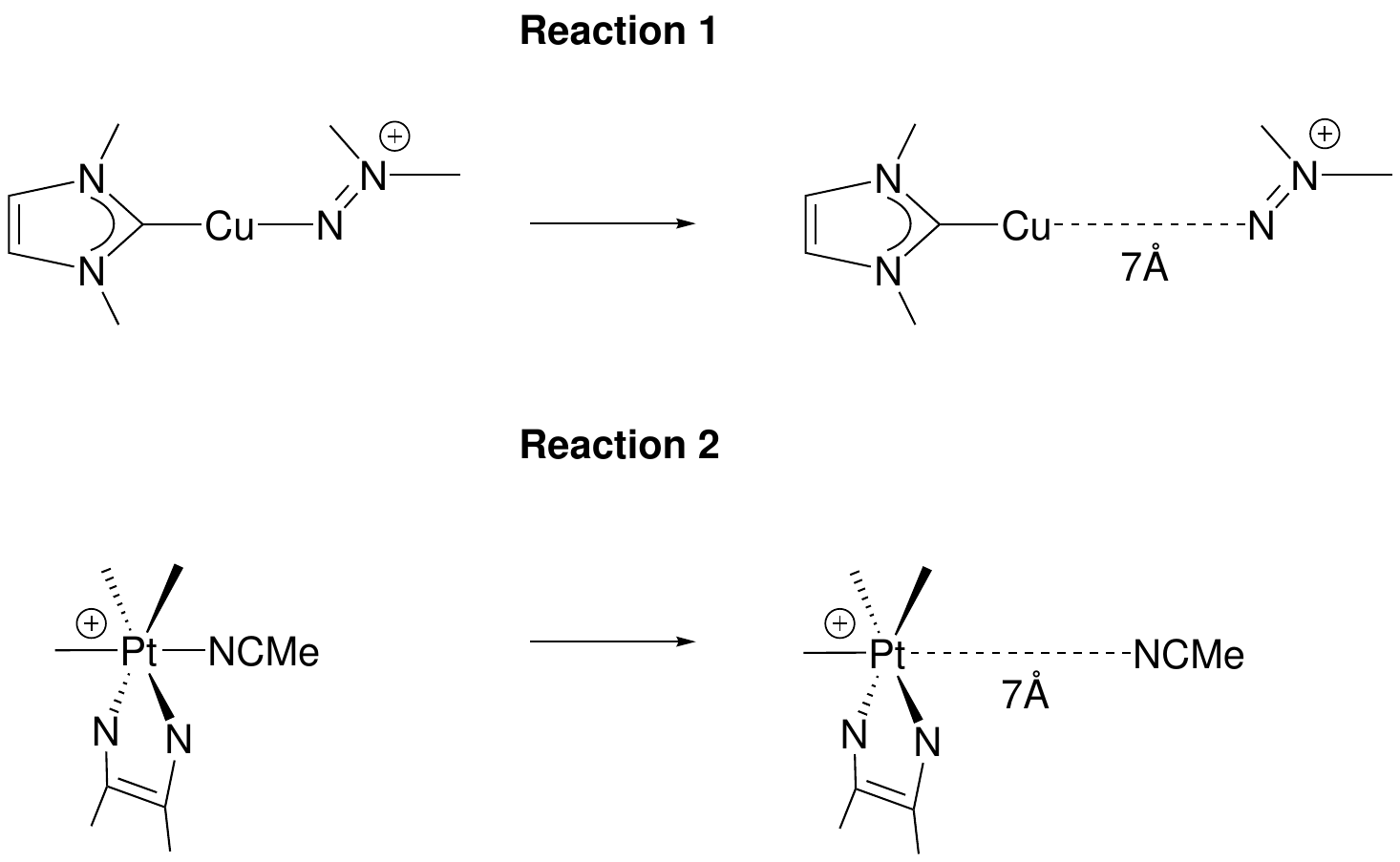}
  \caption{Ligand-binding energies defined by ligand dissociation reactions for two model reactions out of the WCCR10 set of ligand-binding energies.\label{reactions}}
 \end{figure}

For the ligand-dissociation reactions, we applied the structural models depicted in Figure \ref{reactions},
which were truncated compared to the original metal complexes in our previous work \cite{weymuth2014,weymuth2015} on the WCCR10 benchmark set (essentially, large mesityl and 
aromatic groups were replaced by methyl residues). 
The structures were optimized with the BP86 functional\cite{becke1988} and a def2-TZVP\cite{weigend2005} basis set 
along with the corresponding basis set for Coulomb fitting\cite{weigend2006}. 
After the structure optimization, the Cu-N/Pt-N bonds are stretched to 7 \AA~ in order to mimic the final products in a supermolecular calculation for which
the active space can be chosen in complete analogy to the optimized reactant complex (cf. Figure \ref{reactions}). 
The stretched structures were re-optimized with fixed Cu--N and Pt--N internuclear distances, respectively. 
All these preparatory calculations were carried out with the Turbomole program (version 6.5) \cite{ahlrichs1989}. 
Then, pure DFT calculations were carried out with the PBE functional \cite{perdew1996b} and the def-TZVP basis set\cite{bs940415sha} to understand the
effect of the structural truncation as well as the smaller basis set compared to the original work \cite{weymuth2014}.

In the DMRG and DMRG--srPBE calculations, the active spaces were chosen to encompass both $d$-type orbitals and ligand orbitals important for the 
dissociation (see Figures in the Supporting Information\cite{ESI}). The chosen orbital spaces are in all cases very similar between DMRG and DMRG--srPBE.    
DMRG and DMRG--srPBE  calculations  were then performed with the def-TZVP basis set using the same setup as for \ce{H2O} and \ce{N2}. 
The number of renormalized states was $m=2000$ in all cases.  
We also investigated the dissociation energy of reactions 1 and 2, using a larger number of DMRG sweeps and also lower number of renormalized states.
These tests showed that the setup just described was sufficient to achieve convergence within the given significant digits (see Tables in the Supporting Information\cite{ESI}). 
We observe in general that the number of renormalized states required for a converged result is smaller for DMRG--srPBE than for regular DMRG calculations. 
For the two dissociation reactions we have also investigated the effect of using a different short-range DFT functional, namely the Goll--Werner--Stoll\cite{goll2005} combined 
correlation and exchange functional. This combination will be denoted srPBE(GWS). 

All DMRG and DMRG--srDFT calculations in this paper were carried out starting from HF and HF-srDFT orbitals, respectively. 
During the warm-up sweep the state corresponding to the HF determinant was explicitly encoded in the MPS.


\section{\label{results}Results and Discussion}

\subsection{The effect of truncating the active orbital space}\label{sect:active_space_truncation}

The first part of this section is concerned with the effect of a truncation of the CAS both in standard DMRG and DMRG--srDFT calculations on \ce{H2O} and \ce{N2}. 
Total electronic energies are reported and discussed in the Supporting Information\cite{ESI}. While these energies show that the effect of truncation of the active space 
is smaller in DMRG--srPBE than in standard DMRG calculations (due to a 'regularizing' effect of the srDFT part on the CAS), 
the explicit energy data is not meaningful as our current implementation does not support spin-unrestricted
srDFT calculations. As a consequence, the energies of the open-shell reaction products will be asymptotically unreliable (and in fact, worse than those obtained from
unrestricted Kohn--Sham DFT calculations). Moreover, the small active spaces chosen for these two molecules are already so large that good agreement with the 
FCI reference is obtained in the large-CAS DMRG calculations. Therefore, we postpone a discussion of energies to the next subsection. Here, we continue
to explore the 'regularization' effect of srDFT on the active space of the DMRG calculations by investigating the entanglement measures.

 \begin{figure*}[htb!]
  \centering
  \includegraphics[height=18.0cm]{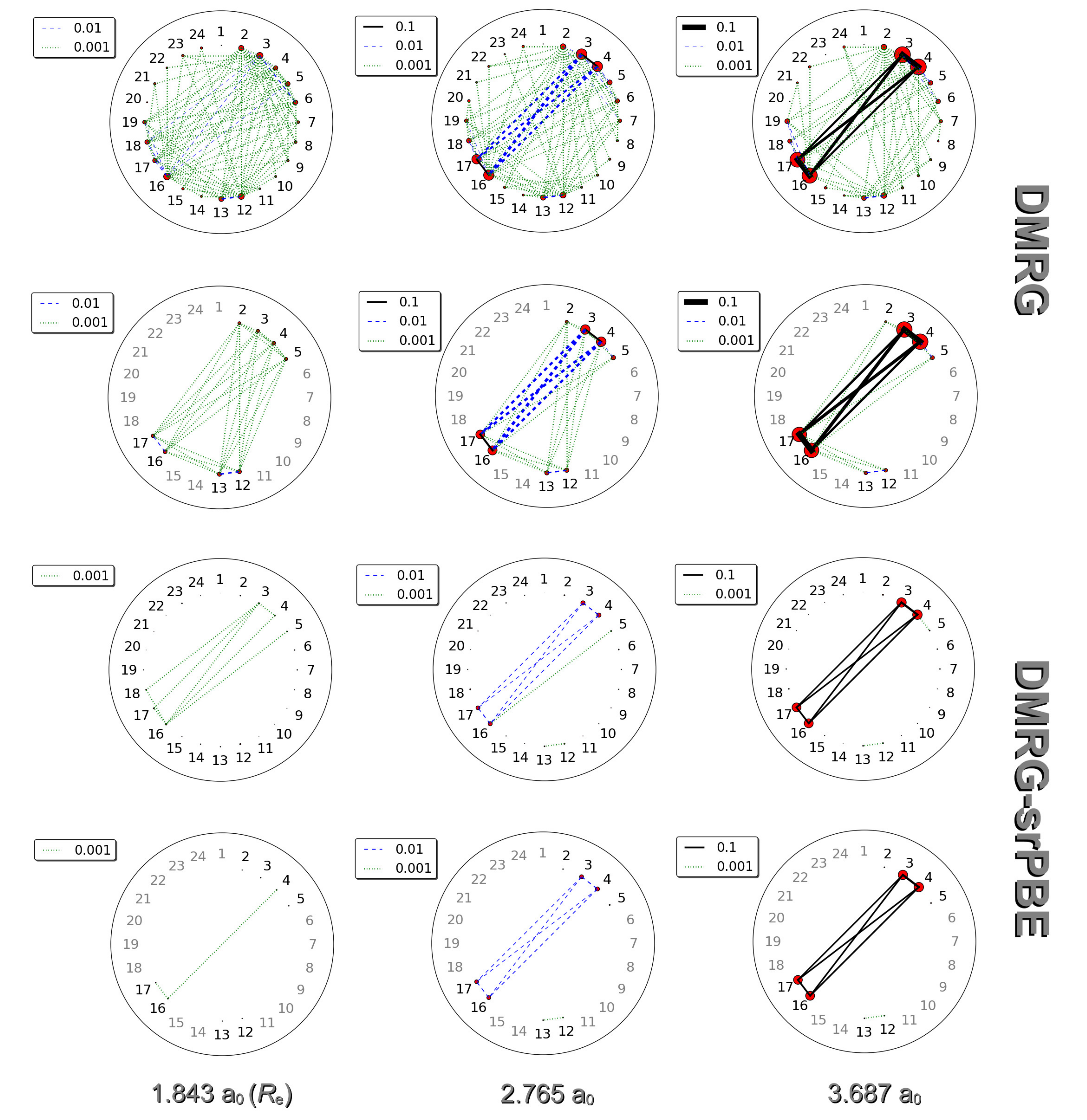}
  \caption{Single-orbital entropies (encoded in the size of the red circles) and mutual information (encoded in the color and strength of the connecting lines)  for \ce{H2O} at three bond distances. 
 Top: Entanglement plots from  DMRG(10,24)[512] with DMRG(8,8)[512] given below.  
Bottom: Entanglement plots from DMRG(10,24)[512]--srPBE with DMRG(8,8)[512]--srPBE given below. 
Active orbital labels are in black and inactive/secondary orbital labels are in gray. The numbering of orbitals is the same for large and small active spaces. 
\label{h2o-entropy}}
 \end{figure*}

\begin{figure*}[htb!]
 \centering
 \includegraphics[height=18.0cm]{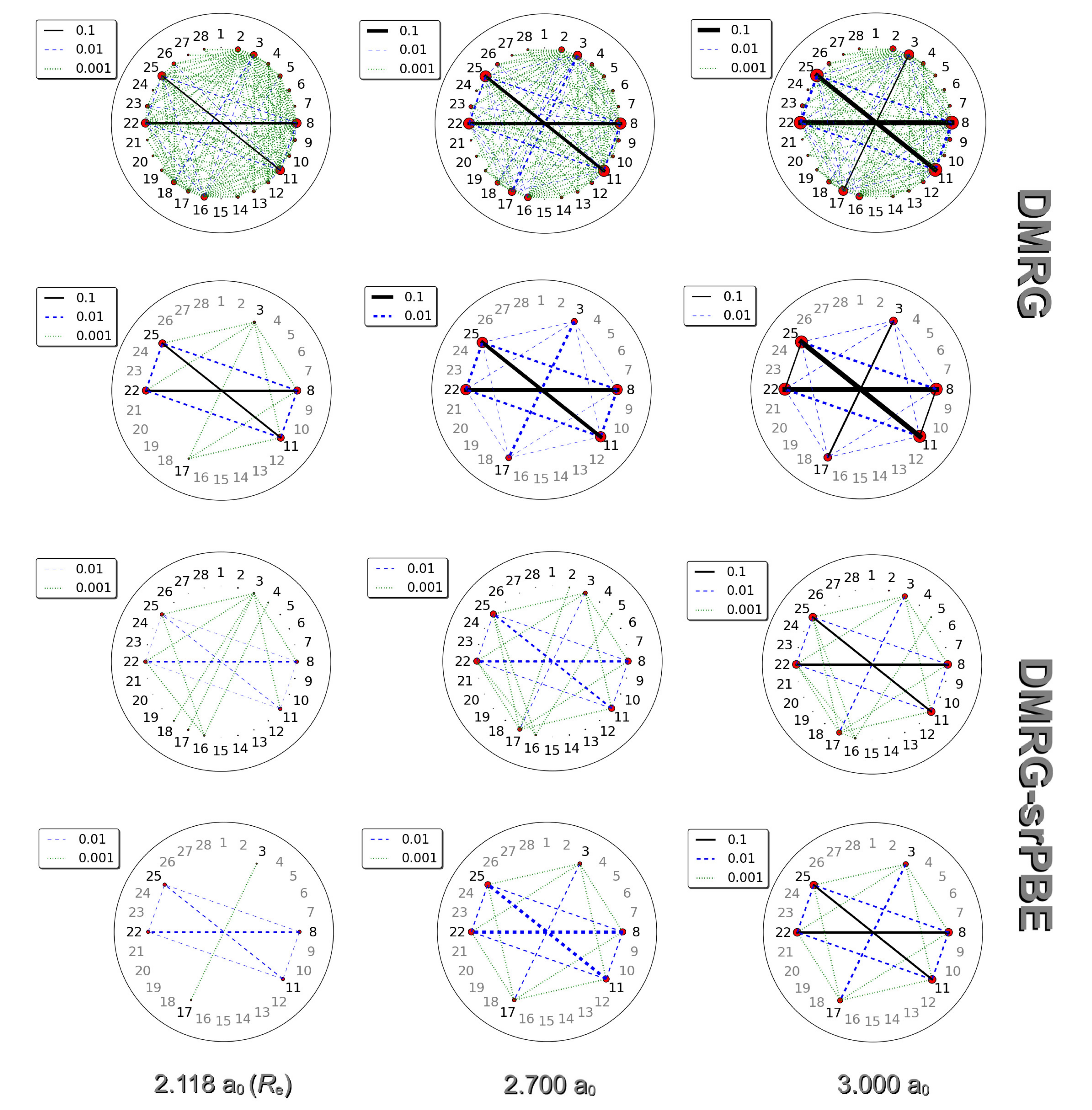}
 \caption{Single-orbital entropies (encoded in the size of the red circles) and mutual information (encoded in the color and strength of the connecting lines) for \ce{N2} at three bond distances. 
Top: Entanglement plots from  DMRG(14,28)[2048] with DMRG(6,6)[2048] given below. Bottom: Entanglement plots from DMRG(14,28)[1024]--srPBE 
with DMRG(6,6)[1024]--srPBE given below. 
Active orbital labels are in black and inactive/secondary orbital labels are in gray. The numbering of orbitals is the same for large and small active spaces. 
\label{n2-entropy}}
\end{figure*}  

In a recent study\cite{boguslawski2012b}, we suggested that single-orbital entropies, $s_u$, and mutual information, $I_{uv}$,
can serve as descriptors to trace and classify multireference character. We found that a balanced active space should 
comprise all orbitals with 
$s_u  > 0.5$ and/or all orbitals within a range of $0.01 < I_{uv} < 0.1$. As these descriptors can be a useful guide to construct optimal active orbital spaces for 
DMRG and standard CAS-type calculations\cite{boguslawski2012b}, they may complement a selection procedure based on natural orbital occupation numbers. 
The single-orbital entropies and mutual information for \ce{H2O} and \ce{N2} calculated at various stretched structures are summarized in 
Figures \ref{h2o-entropy} and \ref{n2-entropy}, respectively. 
In these entanglement plots the magnitude of the single-orbital entropy for each orbital is determined by the size of the corresponding red  circle while the mutual information is encoded by line color and thickness; the thicker and darker the connecting line between two orbitals the larger is their mutual information.   
Considering first the standard DMRG data in the upper parts of Figures \ref{h2o-entropy}  and \ref{n2-entropy},
our entanglement data confirm the chemically intuitive choice that all orbitals involved in \ce{O-H} or \ce{N-N} bonding need to be included in a minimal CAS. 
In addition, in accordance with the boundaries of the entanglement measures defined above for standard DMRG\cite{boguslawski2012b}, the stretched \ce{H2O} and 
\ce{N2} molecules 
display significant multireference character. 
In contrast, both the single-orbital entropies and the mutual information are significantly smaller for DMRG--srDFT 
than for DMRG. For both molecules a major part of the dynamical correlation is indeed treated by the srDFT functional, while static correlation is 
efficiently taken care of by the long-range wave function. Upon bond stretching, static correlation becomes more dominant, but the amount of dynamic correlation 
assigned to the long-range wave function remains effectively constant (cf. the patterns of green lines for the DMRG--srPBE entries in Figures \ref{h2o-entropy} and \ref{n2-entropy} 
when proceeding from left to right). 

Figures  \ref{h2o-entropy} and \ref{n2-entropy} further show the effect of truncating the active spaces. For regular DMRG the 
static correlation increases and appears to be overestimated in the elongated systems for the truncated active spaces. Hence, the effect of active-space truncation 
is much smaller in DMRG--srPBE. Therefore, srDFT has a regularizing effect on the entanglement of orbitals in the active space as the qualitative picture provided by
the entanglement measures does hardly change in DMRG--srDFT calculations when the active space is reduced.

From the above discussion it is also clear that the recommended boundaries for assessing a minimum active orbital space with entanglement measures need to be revised 
for range-separated hybrid methods. Similar conclusions have been drawn with respect to the boundaries for natural occupation numbers as active-orbital space measure 
for the  CASSCF--srDFT\cite{hedegaard2013b} hybrid approach.

\subsection{\label{diss} Ligand binding energies in transition metal complexes}

The ligand-binding energies of the WCCR10 set of ligand dissociation reactions are hard to reproduce by DFT and the PBE functional
turned out to be the pure density functional with the smallest overall error \cite{weymuth2014}. For this reason, we chose two reactions from this test set
to investigate the potential of our DMRG-srDFT approach and employed the structural models depicted in Figure \ref{reactions}.
Tables \ref{reaction_1} and  \ref{reaction_2} provide all data obtained for these two reactions.
We will focus on the pure electronic contribution to the dissociation energy, i.e.\ to $D_e$, as the $D_0$ data are all obtained by the same constant shift and
are reported only as they represent the true experimental observables at zero Kelvin.

First of all, we should discuss the effect that the reduced structural model and the reduced basis-set size have on the dissociation energies reported in Tables \ref{reaction_1} and  \ref{reaction_2}.
This assessment can be done based on the PBE results. As can be seen from the Tables, for reaction 1 we increase the electronic contribution to the
dissociation energy $D_e$ from 247.5 kJ/mol to 257.5 kJ/mol by switching from the quadruple-zeta basis set of Ref.~\citenum{weymuth2014} to triple-zeta basis set employed in this work.
This value is then reduced to 240.2 kJ/mol by reducing the size of structural model. Hence, when considering a zero-point vibrational energy correction (see Table \ref{reaction_1})
the experimental reference result of 218.2 is enlarged to 226.7 to obtain the $D_e$ reference result, which is to be corrected for the reduced basis set by 257.5$-$247.5=+10.0 kJ/mol
and then for the reduced structural model by 240.2$-$257.5=$-$17.3 kJ/mol. The final adjusted reference value for $D_e$ is obtained as 226.7+10.0$-$17.3= 219.4 kJ/mol. 
In a similar manner, we  can adjust the $D_e$ reference energy of 109.9 kJ/mol in Table \ref{reaction_2} by +7.5 kJ/mol for the model-structure error and +12.6 for 
the basis-set error to finally yield 109.9+7.5+12.6=130.0 kJ/mol. 

In Table \ref{reaction_1} for reaction 1, we note that the dissociation energy strongly depends on the size of the active space in the pure DMRG calculations; the dissociation
energy is increased from 132.8 kJ/mol for the small CAS to 173.5 kJ/mol for the largest CAS. This dramatic spread is not seen in the DMRG-srDFT calculations, where it ranges
only from 216.5 kJ/mol to 225.1 kJ/mol. Moreover, we note that these latter energies are in excellent agreement with the reference energy of 219.4 kJ/mol.
Hence, the dynamic correlation captured in the srDFT part allows us to apply a much smaller active space and yields results in much better
agreement with the reference energy. The agreement is also better than the one obtained within pure DFT calculations.
Table \ref{reaction_1} also allows for a comparison of different srDFT functionals. From the dissociation energy of 246.5 obtained with DMRG[2000](30,22)-srPBE(GWS)
and compared to 225.1 obtained with DMRG[2000](30,22)-srPBE we understand that the effect of the approximate functional can be larger than 20 kJ/mol and thus further away from
the reference result (but comparable with the pure DFT result).
 
\begin{table*}[t!]
\centering
\caption{Calculated dissociation energies in kJ/mol for reaction 1 obtained with the def-TZVP basis set. $D_0$ is the zero-point vibrational-energy corrected employing a value of 8.5 kJ/mol for the zero-point vibrational energy obtained for 
the full complex with DFT(BP86)/def2-QZVPP (Refs.~\citenum{weymuth2014,weymuth2015}).   \label{reaction_1} }
\begin{tabular*}{\textwidth}{l@{\extracolsep{\fill}}cc}
\hline \hline \\[-1.5ex]
Method                           & $D_\text{e}$ (kJ/mol) & $D_0$ (kJ/mol) \\[0.5ex]
\hline \\[-1.0ex ]
DMRG[2000](30,22)               &  173.5  & 165.1               \\[0.5ex]
DMRG[2000](20,18)               &  169.9  & 161.5               \\[0.5ex]
DMRG[2000](10,10)               &  132.8  & 124.3               \\[0.5ex]
\hline \\[-1.0ex ]
DMRG[2000](30,22)-srPBE(GWS)    & 246.5   & 238.1              \\[0.5ex]
DMRG[2000](30,22)-srPBE         &  225.1  & 216.6              \\[0.5ex]
DMRG[2000](20,18)-srPBE         &  227.9  &  219.4             \\[0.5ex]
DMRG[2000](10,10)-srPBE         &  216.5  &  208.0             \\[0.5ex]
\hline \\[-1.0ex ]
PBE                             &  240.2  &  231.8           \\[0.5ex] 
PBE (full complex/def2-TZVP)    &  257.5  &  249.0            \\[0.5ex]
PBE (full complex/def2-QZVPP from Ref.~\citenum{weymuth2014}) &  247.5  &  239.0                   \\[0.5ex]
\hline \\[-1.0ex ]
Exp. (Ref.~\citenum{weymuth2014}) &  226.7  &  218.2            \\[0.5ex] 
\hline \hline
 \end{tabular*}
\end{table*}

For reaction 2, we found an even more pronounced dependence of the dissociation 
energy on the size of the active space for the pure DMRG data reported in Table \ref{reaction_2}; 
compare, for instance, 65.3 kJ/mol for DMRG(8,8) to 34.0 kJ/mol for DMRG(22,20).
As this change in energy also increases the deviation from the experimental reference energy although the CAS was enlarged and should have improved on the small-CAS result, 
it calls for a systematic investigation (data reported also in Table \ref{reaction_2}; see also the total electronic energies reported in the Supporting Information\cite{ESI},
which clearly show that the change in dissociation energy is brought about by the product structure only).
We see that the dissociation energies obtained for  the small- to medium-sized active spaces  
in Table \ref{reaction_2} only change moderately. However, for DMRG(18,18) we observe a significant lowering of the dissociation energy, which 
can be explained by closer investigation of the orbitals that were absent in the smaller-CAS calculations (for orbital diagrams and natural occupation numbers 
see the Supporting Information\cite{ESI}): In the product structure, the newly added
 occupied orbital (orbital no.\ 56) differs with an occupation number of 1.966 significantly from 2.000. It is also different 
from the occupation number of 1.992 obtained for its corresponding orbital in the reactant structure (orbital no.\ 61). This large change demands the inclusion of 
the orbital pair for a correct description of the reaction, especially when dynamical correlation is not accounted for otherwise. Notably, the DMRG(8,8) 
to DMRG(16,16) calculations do not include this orbital in the CAS of the product structure, and the better correspondence with experiment must therefore be considered fortuitous. In the 
largest CAS considered, we obtained 44.3 kJ/mol for the electronic contribution to the dissociation energy. Compared to the 34.0 kJ/mol for DMRG(22,20) we thus again start to 
improve the correspondence to the experimental dissociation energy, although the deviation is still dramatic, which might be taken as an indication that 
dynamic correlation is pivotal in the dissociated fragments.

We carried out a similar investigation of the orbitals for the DMRG-srPBE calculations. In this case, the effective inclusion of dynamical correlation 
renders the change in occupation numbers about an order of magnitude smaller. For the orbital pair discussed above, the occupation numbers change 
from 1.996  (orbital no.\ 59) in the reactant to  
1.991 (orbital no.\ 55) in the product. Accordingly, the dissociation energy is less affected (the results being 
80.6 kJ/mol for the smallest CAS and 81.7 kJ/mol for the largest CAS). We note that these results are similar to the corresponding PBE result of 86.4 kJ/mol, but still largely
deviate from the reference value of 130.0 kJ/mol. In order to understand whether the approximate nature of the short-range functional can account for this deviation,
we investigated the alternative srPBE(GWS) functional which yields an dissociation energy of 
101.8 kJ/mol that is considerably closer the reference. Clearly, changing the srDFT functional is not a universal solution as the srPBE(GWS) functional 
for the dissociation energy of reaction 1 yields 246.5 kJ/mol, thus overestimating reaction-1 reference value of 219.4 kJ/mol. Hence, this
srDFT-functional study emphasizes that the available short-range functionals should be improved, as has also been noted by others\cite{fromager2010}.

\begin{table*}[t!]
\centering
\caption{Dissociation energies in kJ/mol for reaction 2 obtained with def-TZVP basis set. 
 $D_0$ is the zero-point vibrational-energy corrected result employing a value of 7.3 kJ/mol for the zero-point vibrational energy obtained for 
the full complex with DFT(BP86)/def2-QZVPP (Ref.~\citenum{weymuth2014,weymuth2015}). 
\label{reaction_2} }
\begin{tabular*}{\textwidth}{l@{\extracolsep{\fill}}cc}
\hline \hline \\[-1.5ex]
Method                                & $D_\text{e}$ (kJ/mol) & $D_0$ (kJ/mol) \\[0.5ex]
\hline \\[-1.0ex ]
DMRG[2000](24,24)                     & 44.3                  & 36.9             \\[0.5ex]
DMRG[2000](22,20)                     & 34.0                  & 26.7             \\[0.5ex]
DMRG[2000](18,18)                     & 37.6	              & 30.2             \\[0.5ex]
DMRG[2000](16,16)                     & 56.6                  & 49.3             \\[0.5ex]
DMRG[2000](14,14)                     & 55.2                  & 47.8             \\[0.5ex]
DMRG[2000](10,10)                     & 60.4                  & 53.1             \\[0.5ex]
DMRG[2000](8,8)                       & 65.3                  & 58.0             \\[0.5ex]
\hline \\[-1.0ex ]                     
DMRG[2000](22,20)-srPBE(GWS)          & 101.8                 & 94.5           \\[0.5ex]
DMRG[2000](22,20)-srPBE               & 81.7                  & 74.3           \\[0.5ex] 
DMRG[2000](8,8)-srPBE                 & 80.6                  & 73.2            \\[0.5ex] 
\hline \\[-1.0ex ]             
PBE                                   & 86.4                  & 79.1           \\[0.5ex]
PBE (full complex/def2-TZVP)          & 73.8                  & 66.5               \\[0.5ex]
PBE (full complex/def2-QZVPP from Ref.~\citenum{weymuth2014})&  66.3  & 59.0    \\[0.5ex]
\hline \\[-1.0ex ]
Exp. (Ref.~\citenum{weymuth2014})     & 109.9     &  102.6    \\[0.5ex]
\hline \hline
 \end{tabular*}
\end{table*}

\section{\label{conclusion} Conclusions and outlook}

We have presented the development and first implementation of a hybrid approach that couples DMRG and DFT using a range-separation \emph{ansatz}. 
To analyze the DMRG--srDFT approach, we considered the (symmetrically) dissociating \ce{H2O} and \ce{N2} as well as ligand-binding energies in transition-metal complexes.

Although total electronic energies from small-CAS DMRG calculations were found to be very close to the corresponding FCI results for the two small molecules, we found that 
the effect of truncation of the active space is smaller for DMRG--srPBE than for standard DMRG calculations. This effect was also visible in the entanglement-entropy measures considered.
We studied the effect of the simultaneous treatment of static and dynamic correlation on such orbital-based descriptors, namely on single-orbital entropies and mutual information. 
We find that (i) the single-orbital entropies and mutual information are consistently smaller for DMRG--srDFT than for DMRG and that (ii)
the major part of dynamical correlation is assigned to the short-range DFT part so that the pattern of these entropy measures hardly change when the
active space is reduced in a DMRG-srDFT calculation.

The discussion of ligand-binding energies revealed the true potential of the DMRG--srDFT approach. 
For two reactions out of the WCCR10 benchmark set of 
ligand-binding energies, we found that DMRG-srDFT yields a much less pronounced dependence of the reaction energy on the size of the active space such that
more consistent results were obtained. Focusing on one functional (PBE and its short-range variant), we found for reaction 1 a significant improvement on
both pure DMRG and pure PBE results. However, the situation was more delicate in the case of reaction 2 for which we found a dramatic dependence of the
reaction energy on the inclusion of specific orbitals in the active space such that the pure DMRG result deviates significantly from the experimental reference,
indicating a very pronounced contribution of orbitals beyond the chosen active spaces to the correlation energy. Still, the final DMRG-srPBE result turned out to be very 
similar to the pure PBE result. However, we also noted
that all data still deviated much from the experimental reference, which might even be taken as an indication to reinvestigate the accuracy of the experimental value.

With DMRG as the wave function part, DMRG--srDFT provides access to much larger complete active orbital spaces than those feasible with any traditional CAS-type approach combined with DFT. 
Hence, including dynamical correlation through a short-range density functional is a viable
option to preserve the efficiency of DMRG calculations by avoiding standard perturbation-theory-based approaches.
This facilitates calculations with long-range CAS-type wave functions such that all remaining approximations are buried in the approximate short-range DFT functional. 

Although we obtained encouraging results for our case studies when comparing our new DMRG--srDFT approach with truncated active orbital 
spaces to standard DMRG and FCI, further improvement is possible and work along these lines is in progress in our laboratory.
Apart from an extension of our implementation to a spin-unrestricted framework in the srDFT part, also spin-state-specific short-range functionals will further be crucial for molecules with an open-shell electronic structure \cite{jacob2012}.

Alternative approaches for a simultaneous treatment of static and dynamic correlation in a hybrid DMRG approach that avoid 
a range-separation \emph{ansatz} exist. In future work, we are considering the pair-density functional theory which was recently put forward for MCSCF 
methods\cite{manni2014} as well as the \textit{site occupation} density functional theory proposed by Fromager \cite{fromager2015}. 

It should finally be emphasized that hybrids between DFT and wave function methods are also 
expected to have less dramatic dependence on the one-electron basis set than standard wave function methods. 
We are currently investigating this more quantitatively.     

\begin{acknowledgments}

EDH thanks the Villum foundation for a post-doctoral fellowship.
This work has been financially supported by ETH Zurich and the Schweizer Nationalfonds (SNF project 200021L\_156598).
HJAaJ thanks the Danish Council for Independent Research $|$ Natural Sciences for financial support (grant number 12-127074).

\end{acknowledgments}


\clearpage
\newpage


\section*{Supplementary material}

**Electronic Supporting Infomation:**

**Density Matrix Renormalization Group with Efficient Dynamical Electron Correlation Through Range Separation**

Erik Donovan Hedegård,[1, a)] Stefan Knecht,[1] Jesper Skau Kielberg,[2] Hans Jørgen Aagaard Jensen,[2, b)] and Markus Reiher[1, c)]

[1)] *ETH Zürich, Laboratorium für Physikalische Chemie, Vladimir-Prelog Weg-2, CH-8093 Zürich, Switzerland*

[2)] *Department of Physics, Chemistry and Pharmacy, University of Southern Denmark, Campusvej 55, Odense, Denmark*

(Dated: May 27, 2015)

———

[a)] Electronic mail: erik.hedegard@phys.chem.ethz.ch
[b)] Electronic mail: hjj@sdu.dk
[c)] Electronic mail: markus.reiher@phys.chem.ethz.ch

# I. TOTAL ELECTRONIC ENERGIES FOR WATER AND DINITROGEN

$H_2O$ and $N_2$ have been investigated extensively with both traditional wave function methods and DMRG[1–3]. Here, we compare to the work by Olsen *et al.*[4] for $H_2O$ and the studies by Krogh and Olsen[5], Larsen *et al.*[6], and Chan *et al.*[2] for $N_2$.

In the following, the DMRG and DMRG–srPBE results are compared in terms of the total energy and of the non-parallelity error (NPE). The NPE measures the difference between the absolute maximum and minimum deviations from the DMRG-FCI potential curve over the internuclear distances considered (see e.g. Ref. 7). The NPE would be zero if the curves differed by a constant shift. As the NPE suffers from that the individual methods do not share the same energy scale (due to the approximate srDFT functional) we have additionally reported the energy difference (in kJ mol$^{-1}$) with respect to the electronic equilibrium internuclear distance ($R_e$).

Starting with $H_2O$, the energies calculated at several distances along the symmetric stretching coordinate are summarized in Table I. The DMRG(10,24)[512] data agree well with the FCI results by Olsen *et al.*[4] which also shows that the number of renormalized active-subsystem block states $m = 512$ is sufficient to achieve FCI accuracy with DMRG. It is further seen that the DMRG–srPBE energies are higher than the energies of DMRG(10,24)[512], even for DMRG(10,24)[512]–srPBE where a FCI long-range wave function is employed. Although one might intuitively expect that additional dynamical correlation would lower the energy, the approximate nature of the srDFT functionals should again not be forgotten. As mentioned in the Theory section of the main article, there is no guarantee that the WFT–srDFT energy for a given $\mu$-value is below the corresponding WFT energy. A similar result has been observed by Fromager *et al.*[8]

Finally, we should note that the DMRG(10,24)[512]–srPBE data can be considered as a technical consistency test of the approach. A CAS(10,24) active space represents the full configuration interaction scenario, for which the srDFT should not give any additional contribution within the one-electron basis set chosen. For $H_2O$ we understand from the DMRG(10,24)[512]–srPBE result in Table I that $918 - 762 = 156$ kJ/mol are artificially introduced due to two technical limitations: (i) the spin-restricted implementation of the srDFT part and (ii) the approximate nature of the short-range functional.

Turning to $N_2$, all results are compiled in Table II. Our reference DMRG(14,28)[2048]

results are in good agreement with the results reported by Chan and co-workers[2], although small differences occur far from the equilibrium distance, presumably due to different procedures during the DMRG start-up sweeps.

For both $H_2O$ and $N_2$, the effect of truncation of the active space is smaller for DMRG–srPBE than for standard DMRG. In the main paper, we have elaborated on this issue by analysis of the entanglement entropies.

With respect to the NPEs of DMRG–srPBE, Tables I and II take the standard FCI data as reference. For this reference, the NPEs are for $H_2O$ 0.0595 (DMRG(10,24)[512]–srPBE) and 0.0592 (DMRG(8,8)[512]–srPBE), respectively. Thus, for $H_2O$, the NPE of DMRG–srPBE is in between the NPEs of PBE (0.0342) and PBE0 (0.0677). The NPEs are for $N_2$ 0.0986 (DMRG(14,28)[1024]–srPBE) and 0.0984 (DMRG(6,6)[1024]–srPBE), whereas the PBE functional gives an NPE of 0.0691. We note that for $N_2$, the PBE0 functional even crosses the DMRG(14,28)[2048] curve, which is neither observed for DMRG–srPBE nor for standard PBE, and which makes it impossible to calculate a well-defined NPE. More importantly, the NPEs of the truncated-active-space DMRG–srPBE are in both cases, $H_2O$ and $N_2$, nearly identical up to $\Delta$(NPE) = 0.0003 with their DMRG(FCI)-srPBE references. *This substantiates the above finding that even for truncated active spaces accurate results close to DMRG(FCI)–srDFT quality are within reach; an observation that will be important for larger molecules where FCI is beyond the limits, even for DMRG.*

It should also be mentioned that the NPEs for both $H_2O$ and $N_2$ with DMRG–srDFT are larger than for pure DMRG, also when using a DMRG(FCI)–srDFT description. The same holds for the energies relative to $R_e$, which are overestimated (this is also the case for regular DFT). Since one for DMRG(FCI)-srDFT have an exact description of the long-range wave function, the errors arise from the approximate srDFT functional. One obvious source of error is that the used functionals are not spin-adapted. Further, the applied short-range functional is not self-interaction free, as discussed in Ref. 8.

From the discussion of the total electronic energies we can conclude that:

- The small active space is already so large for these small molecules that dynamical correlation plays a minor role. Consequently, relative energies are already very well reproduced by the small CAS.

- Still, it can be seen that the effect of truncation of the active space is smaller for

TABLE I. DMRG, DMRG–srDFT, FCI, and DFT energies for H$_2$O at various O−H distances along the symmetric streching mode in a Dunning cc-pVDZ basis set. Values in parantheses are differences (in kJ mol$^{-1}$) with respect to the electronic equilibrium internuclear distance ($R_e$) for a given setup. DMRG-srDFT data at increasing distance become unreliable as our implementation is a spin-restructed one that cannot properly describe the open-shell fragments. a$_0$ denotes the Bohr radius.

| Method | 1.843 a$_0$ ($R_e$) | 2.765 a$_0$ | 3.687 a$_0$ | NPE |
|---|---|---|---|---|
| FCI = CAS(10,24) (Ref. 4) | −76.241 860 (0.0) | −76.072 348 (445) | −75.951 711 (762) | — |
| DMRG(10,24)[512] | −76.241 748 (0.0) | −76.072 206 (445) | −75.951 577 (762) | — |
| DMRG(8,8)[512] | −76.064 887 (0.0) | −75.906 261 (417) | −75.797 043 (703) | 0.0223 |
| DMRG(10,24)[512]–srPBE | −76.334 180 (0.0) | −76.153 803 (474) | −75.984 516 (918) | 0.0595 |
| DMRG(8,8)[512]–srPBE | −76.331 452 (0.0) | −76.150 984 (474) | −75.982 104 (917) | 0.0592 |
| PBE | −76.333 974 (0.0) | −76.173 674 (421) | −76.018 828 (827) | 0.0342 |
| PBE0 | −76.338 765 (0.0) | −76.155 765 (481) | −75.980 886 (940) | 0.0677 |

DMRG–srPBE than for standard DMRG.

- Total electronic energies at increasing internuclear distance and thus relative energies are unreliable because of a technical limitation of our current implementation (that is a restriction to spin-restricted electronic structures, which is not valid in the asymptotic limit).

## II. LIGAND-BINDING ENERGIES

To demonstrate that convergence w.r.t. the DMRG macro-iterations (sweeps) has been achieved, we present the corresponding data in Tables III and IV. Total electronic energies for the ligand-dissociation reactions are reported in Tables V and VI.

TABLE II. DMRG, DMRG–srDFT, and DFT energies for $N_2$ at various N–N distances in a Dunning cc-pVDZ basis set. All energies are in Hartree. Values in parantheses are differences (in kJ mol$^{-1}$) with respect to the electronic equilibrium internuclear distance ($R_e$) for a given setup. DMRG-srDFT data at increasing distance become unreliable as our implementation is a spin-restructed one that cannot properly describe the open-shell fragments. $a_0$ denotes the Bohr radius.

| Method | 2.118 $a_0$ ($R_e$) | 2.4 $a_0$ | 2.7 $a_0$ | 3.0 $a_0$ | 3.6 $a_0$ | NPE |
|---|---|---|---|---|---|---|
| DMRG(14,28)[2048] | −109.281 844 (0) | −109.241 450 (107) | −109.162 965 (314) | −109.088 526 (511) | −108.996 481 (756) | — |
| DMRG(14,28)[1024] | −109.281 050 | −109.240 379 | −109.161 507 | −109.086 592 | −108.993 197 | — |
| DMRG(14,28)[2000] (Ref. 2) | −109.282 088 | −109.241 803 | −109.163 467 | −109.089 252 | −108.997 939 | — |
| DMRG(14,28)[1000] (Ref. 2) | −109.281 877 | −109.241 423 | −109.163 087 | −109.088 772 | −108.997 549 | — |
| DMRG(6,6)[1024] | −109.021 680 (0) | −108.977 348 (116) | −108.902 827 (312) | −108.833 061 (495) | −108.750 501 (712) | 0.0203 |
| DMRG(14,28)[1024]–srPBE | −109.403 134 (0) | −109.350 060 (139) | −109.256 169 (386) | −109.162 722 (631) | −109.016 692 (1015) | 0.0986 |
| DMRG(6,6)[1024]–srPBE | −109.397 043 (0) | −109.344 068 (139) | −109.250 387 (385) | −109.156 976 (630) | −109.010 779 (1014) | 0.0984 |
| PBE | −109.414 21 (0) | −109.370 598 (115) | −109.285 622 (338) | −109.198 659 (566) | −109.057 259 (937) | 0.0691 |
| PBE0 | −109.409 399 (0) | −109.353 294 (147) | −109.254 439 (407) | −109.153 953 (671) | −108.988 792 (1104) | N.A. |

TABLE III. Dissociation energies for reaction 1, obtained with different number of renormalized states ($m$) and sweeps. The results have not been corrected for the zero-point vibrational energy.

| Method | Sweeps | $D_e$ (kJ/mol) |
|---|---|---|
| DMRG[2000](30,22) | 20 | 173.5 |
| DMRG[2000](30,22) | 10 | 173.5 |
| DMRG[1000](30,22) | 10 | 173.5 |
| DMRG[2000](30,22)-srPBE | 20 | 225.1 |
| DMRG[2000](30,22)-srPBE | 10 | 225.1 |
| DMRG[1000](30,22)-srPBE | 10 | 225.1 |

TABLE IV. Dissociation energies for reaction 2, obtained with different number of renormalized states ($m$) and sweeps. The results have not been corrected for the zero-point vibrational energy.

| Method | Sweeps | $D_e$ (kJ/mol) |
|---|---|---|
| DMRG[2000](22,20) | 20 | 34.0 |
| DMRG[2000](22,20) | 10 | 34.0 |
| DMRG[1000](22,20) | 10 | 34.0 |
| DMRG[2000](22,20)-srPBE | 20 | 81.6 |
| DMRG[2000](22,20)-srPBE | 10 | 81.6 |
| DMRG[1000](22,20)-srPBE | 10 | 81.7 |

TABLE V. Total electronic energies in Hartree for ligand-dissociation reaction 1 along with corresponding dissociation energies (def-TZVP basis set). $D_0$ is zero-point vibrational-energy corrected (see the main paper).

| Method | Reactant | Product | $D_e$ (kJ/mol) | $D_0$ (kJ/mol) |
|---|---|---|---|---|
| DMRG[2000](30,22) | -2129.899 308 | -2129.832 571 | 173.5 | 165.1 |
| DMRG[2000](20,18) | -2129.896 908 | -2129.831 567 | 169.9 | 161.5 |
| DMRG[2000](10,10) | -2129.880 839 | -2129.829 791 | 132.8 | 124.3 |
| DMRG[2000](30,22)-srPBE | -2133.524 563 | -2133.438 019 | 225.1 | 216.6 |
| DMRG[2000](20,18)-srPBE | -2133.524 455 | -2133.436 822 | 227.9 | 219.4 |
| DMRG[2000](10,10)-srPBE | -2133.518 977 | -2133.435 733 | 216.5 | 208.0 |
| DMRG[2000](30,22)-srPBE(GWS) | -2133.609 491 | -2133.514 696 | 246.5 | 238.1 |

TABLE VI. Total electronic energies in Hartree for ligand-dissociation reaction 2 along with corresponding dissociation energies (def-TZVP basis set). $D_0$ is zero-point vibrational-energy corrected (see the main paper).

| Method | Reactant | Product | $D_e$ (kJ/mol) | $D_0$ (kJ/mol) |
|---|---|---|---|---|
| DMRG[2000](24,24) | -712.280 859 | -712.263 842 | 44.3 | 36.9 |
| DMRG[2000](22,20) | -712.269 411 | -712.256 329 | 34.0 | 26.7 |
| DMRG[2000](18,18) | -712.265 749 | -712.251 309 | 37.6 | 30.2 |
| DMRG[2000](16,16) | -712.260 395 | -712.238 611 | 56.6 | 49.3 |
| DMRG[2000](14,14) | -712.255 088 | -712.233 875 | 55.2 | 47.8 |
| DMRG[2000](10,10) | -712.251 167 | -712.227 939 | 60.4 | 53.1 |
| DMRG[2000](8,8) | -712.249 671 | -712.224 544 | 65.3 | 58.0 |
| DMRG[2000](22,20)-srPBE | -716.065 249 | -716.0338 492 | 81.7 | 74.3 |
| DMRG[2000](8,8)-srPBE | -716.058 103 | -716.0271 214 | 80.6 | 73.2 |
| DMRG[2000](22,20)-srPBE(GWS) | -716.065 249 | -716.033 849 | 101.8 | 94.5 |

TABLE VII. Natural orbital occupation numbers for reactant and product structures of reaction 2. The orbitals are shown in Figures 5-8.

| Orbital | Reactant | Product | Reactant | Product | Reactant | Product | Reactant | Product |
|---|---|---|---|---|---|---|---|---|
| | DMRG(16,16) | | DMRG(18,18) | | DMRG(22,20) | | DMRG(22,20)-srPBE | |
| 54 | Inactive | Inactive | Inactive | Inactive | 1.99927 | 1.99999. | 1.99975 | 1.99984 |
| 55 | Inactive | Inactive | Inactive | Inactive | 1.99983 | 1.98724 | 1.99982 | 1.99053 |
| 56 | Inactive | Inactive | 1.99965 | 1.96652 | 1.99947 | 1.96575 | 1.99928 | 1.99420 |
| 57 | 1.99248 | 1.99860 | 1.99393 | 1.99857 | 1.99521 | 1.99851 | 1.99986 | 1.99999. |
| 58 | 1.99927 | 1.99892 | 1.99785 | 1.99950 | 1.99665 | 1.99950 | 1.99214 | 1.99621 |
| 59 | 1.99979 | 1.99618 | 1.99958 | 1.99571 | 1.99959 | 1.99868 | 1.99658 | 1.99774 |
| 60 | 1.99874 | 1.97459 | 1.99822 | 1.97403 | 1.99303 | 1.96981 | 1.99984 | 1.99642 |
| 61 | 1.99515 | 1.99124 | 1.99166 | 1.99003 | 1.98939 | 1.98987 | 1.99423 | 1.99999. |
| 62 | 1.97704 | 1.99999. | 1.97567 | 1.99999. | 1.97617 | 1.99999. | 1.99581 | 1.99999. |
| 63 | 1.99611 | 1.99999. | 1.99002 | 1.99999. | 1.99056 | 1.99999. | 1.99914 | 1.99916 |
| 64 | 1.99787 | 1.99224 | 1.99615 | 1.99265 | 1.99614 | 1.99273 | 1.99560 | 1.99987 |
| 65 | 0.02884 | 0.02590 | 0.03046 | 0.02689 | 0.03189 | 0.03606 | 0.01009 | 0.00597 |
| 66 | 0.00005 | 0.00638 | 0.00005 | 0.03751 | 0.00008 | 0.03861 | 0.00675 | 0.01128 |
| 67 | 0.00002 | 0.00001 | 0.00003 | 0.00001 | 0.00006 | 0.00003 | 0.00631 | 0.00646 |
| 68 | 0.00493 | 0.00005 | 0.01314 | 0.00003 | 0.01347 | 0.00004 | 0.00003 | 0.00001 |
| 69 | 0.00761 | 0.00047 | 0.00902 | 0.00071 | 0.00983 | 0.00075 | 0.00066 | 0.00175 |
| 70 | 0.00032 | 0.01288 | 0.00010 | 0.01470 | 0.00013 | 0.01512 | 0.00005 | 0.00045 |
| 71 | 0.00012 | 0.00004 | 0.00017 | 0.00010 | 0.00023 | 0.00013 | 0.00065 | 0.00005 |
| 72 | 0.00165 | 0.00251 | 0.00179 | 0.00295 | 0.00467 | 0.00705 | 0.00333 | 0.00006 |
| 73 | Secondary | Secondary | 0.00250 | 0.00011 | 0.00432 | 0.00012 | 0.00006 | 0.00003 |
| 74 | Secondary | Secondary | Secondary | Secondary | Secondary | Secondary | Secondary | Secondary |

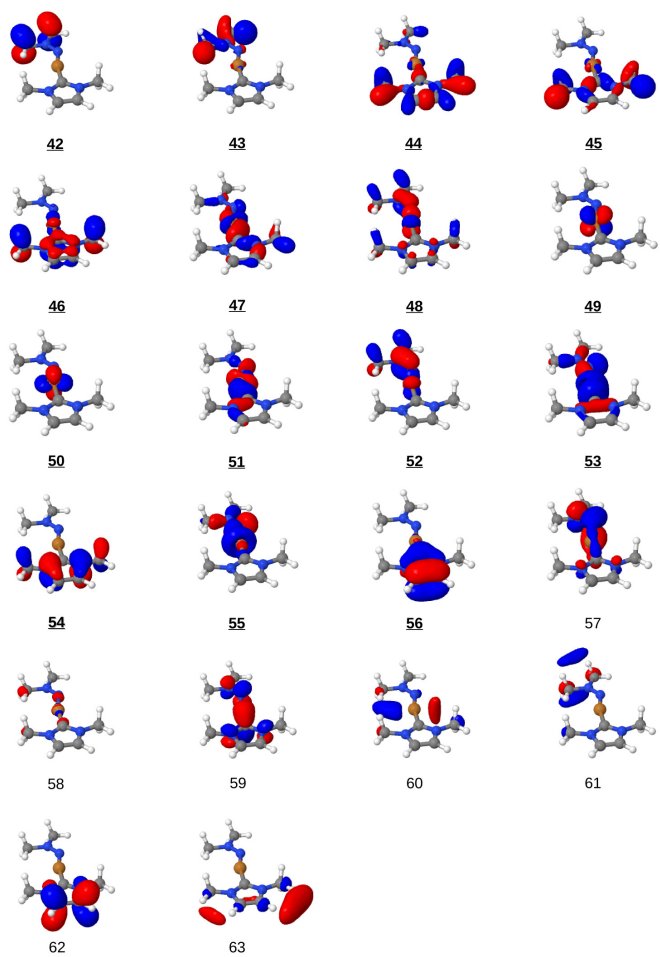

FIG. 1. Active space orbitals for the complex in reaction 1. The occupied orbitals are underlined in bold font.

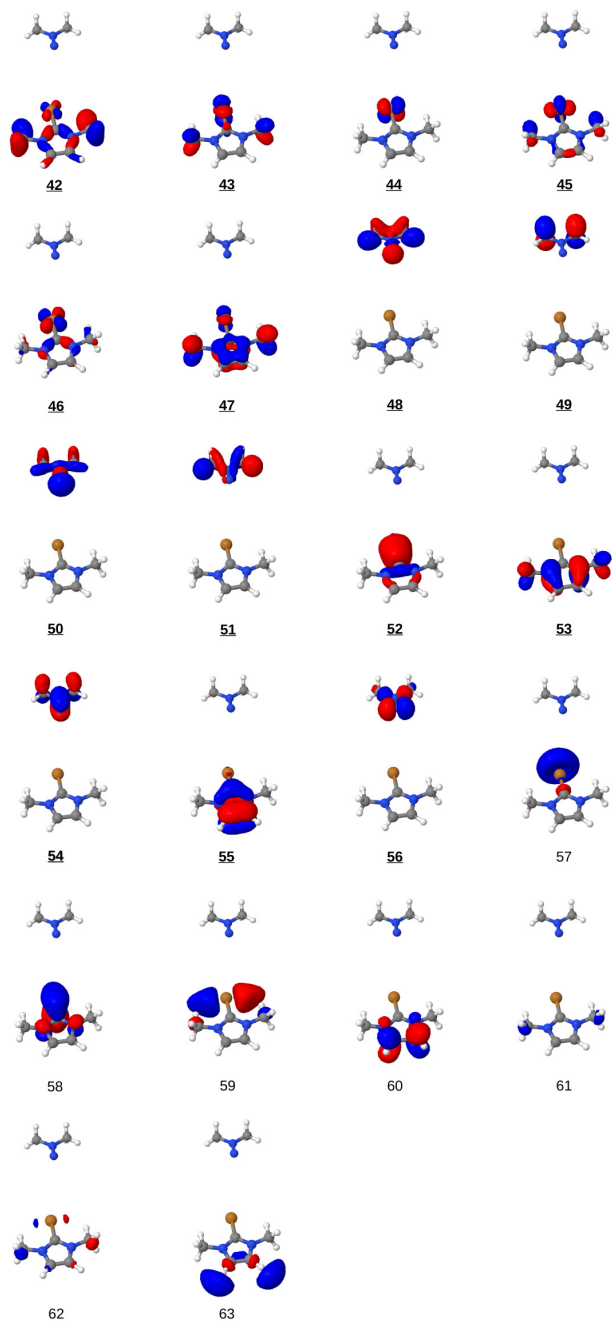

FIG. 2. Active space orbitals for the stretched structure (Cu–N internuclear distance 7 Å) in reaction 1. The occupied orbitals are underlined in bold font.

9

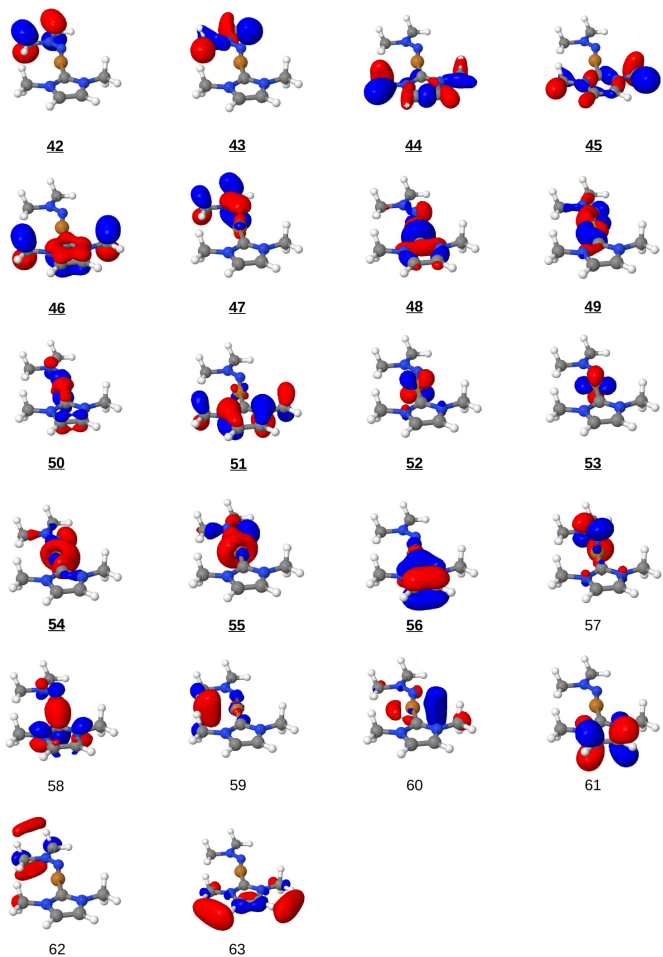

FIG. 3. Active space orbitals for the complex in reaction 1. The occupied orbitals are underlined in bold font.

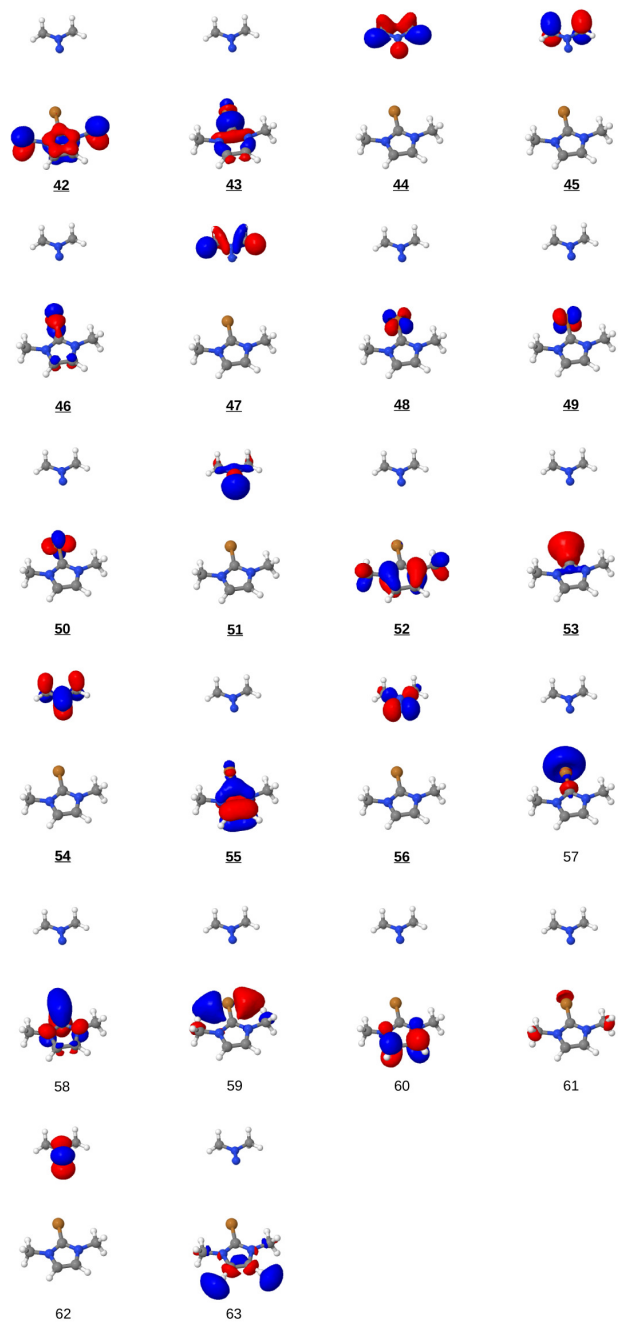

FIG. 4. Active space orbitals for the stretched structure (Cu–N internuclear distance 7 Å) in reaction 1. The occupied orbitals are underlined in bold font.

11

**DMRG[2000](22,20) Active Space Orbitals**

|  |  |  |  |
|---|---|---|---|
| **_54_** | **_55_** | **_56_** | **_57_** |
| **_58_** | **_59_** | **_60_** | **_61_** |
| **_62_** | **_63_** | **_64_** | 65 |
| 66 | 67 | 68 | 69 |
| 70 | 71 | 72 | 73 |

FIG. 5. Active space orbitals for the complex in reaction 2. The occupied orbitals are underlined in bold font.

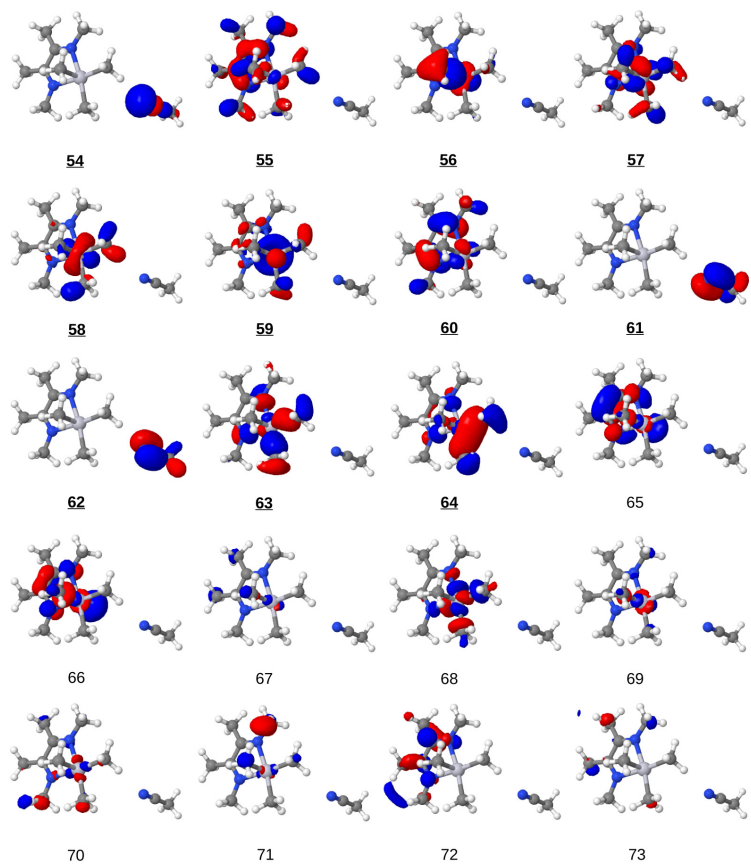

FIG. 6. Active space orbitals for the stretched structure (Pt–N internuclear distance 7 Å) in reaction 2. The occupied orbitals are underlined in bold font.

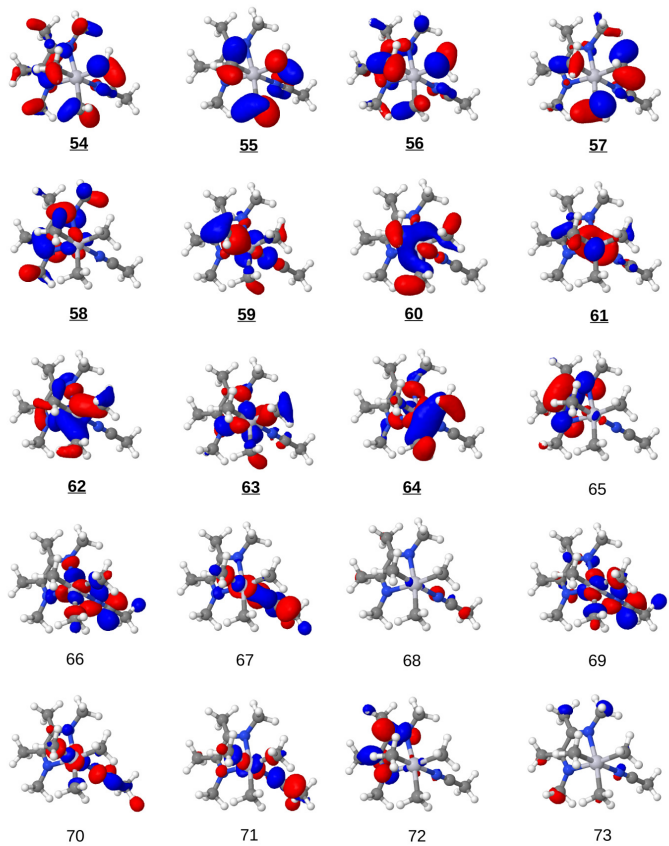

FIG. 7. Active space orbitals for the complex in reaction 2. The occupied orbitals are underlined in bold font.

14

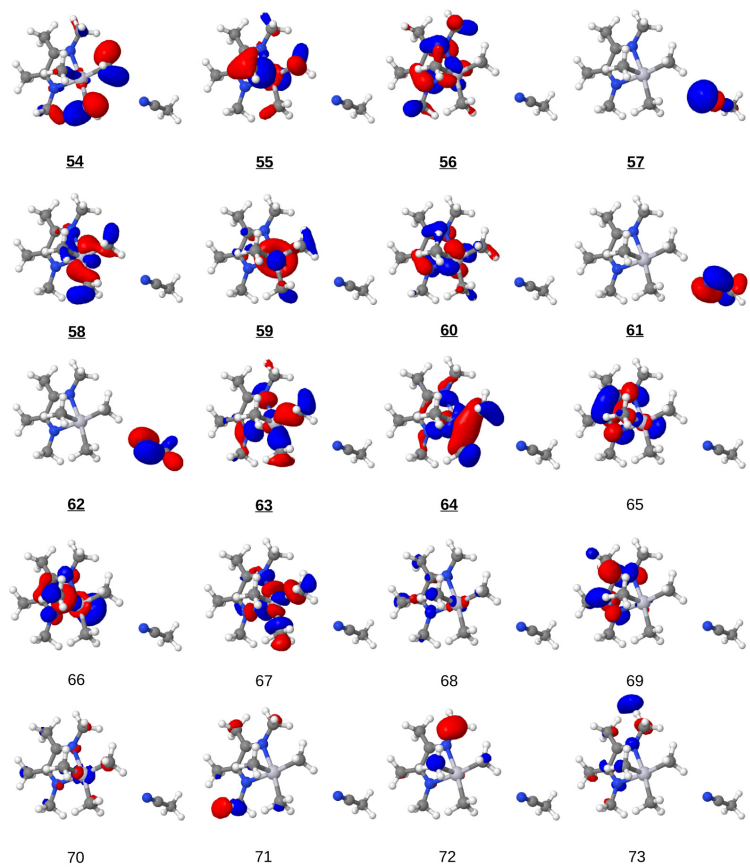

FIG. 8. Active space orbitals for the stretched structure (Pt–N internuclear distance 7 Å) in reaction 2. The occupied orbitals are underlined in bold font.

15



\end{document}